\documentclass[fleqn,usenatbib]{mnras}

\usepackage{newtxtext,newtxmath}

\usepackage[T1]{fontenc}

\DeclareRobustCommand{\VAN}[3]{#2}
\let\VANthebibliography\thebibliography
\def\thebibliography{\DeclareRobustCommand{\VAN}[3]{##3}\VANthebibliography}

\usepackage{graphicx}	
\usepackage{amsmath}	

\usepackage[font=small,justification=centering,labelfont=bf]{caption} 

\usepackage{natbib} 

\title[DART ejecta and dust cloud]{Lofting of low speed ejecta produced in the DART experiment and production of a dust cloud}

\author[G. Tancredi et al.]{
Gonzalo Tancredi,$^{1}$\thanks{E-mail: gonzalo@fisica.edu.uy}
Po-Yen Liu,$^{2}$
Adriano Campo-Bagatin,$^{2}$
Fernando Moreno,$^{3}$
and Bruno Dom\'inguez$^{1}$
\\
$^{1}$Departamento de Astronom\'ia, Facultad de Ciencias, Igu\'a 4225, 11400 Montevideo, Uruguay\\
$^{2}$Departamento de F\'isica, Ingenier\'ia de Sistemas y Teor\'ia de la Se\~nal, Universidad de Alicante, San Vicent del Raspeig, 03690 Alicante, Spain\\
$^{3}$Instituto de Astrof\'isica de Andaluc\'ia, CSIC, Glorieta de la Astronom\'ia s/n, 18008 Granada, Spain
}

\date{Accepted XXX. Received YYY; in original form ZZZ}

\pubyear{2022}

\begin{document}
\label{firstpage}
\pagerange{\pageref{firstpage}--\pageref{lastpage}}
\maketitle

\begin{abstract}
NASA sent the DART (Double Asteroid Redirection Test) mission to impact Dimorphos, the satellite of the asteroid binary system (65803) Didymos. DART will release LICIACube prior to impact to obtain high-resolution post-impact images. The impact will produce a crater and a large amount of material ejected at high speed (several tens of $m/s$), producing an ejecta cone that will quickly disperse.

We analyzed an additional effect: the lofting of material at low velocity due to the generation of seismic waves that propagate inside Dimorphos, producing surface shaking far from the impact point.
We divide the process into different stages: from the generation of impact--induced waves, the interaction of them with surface particles, the ejection of dust particles at velocities, and the prediction of the observability of the dust coma and trail.

We anticipate the following observable effects: \textit{i)} generation of a dust cloud that will produce a hazy appearance of Dimorphos' surface, detectable by LICIACube; \textit{ii)} brightness increase of the binary system due to enhancement of the cross section produced by the dust cloud; \textit{iii)} generation of a dust trail, similar to those observed in some Active Asteroids, which can last for several weeks after impact. 

Numerical prediction of the detectability of these effects depends on the amount and size distribution of ejected particles, which are largely unknown. In case these effects are observable, an inversion method can be applied to compute the amount of ejected material and its velocity distribution, and discuss the relevance of the shaking process.


\end{abstract}




\section{Introduction}

Humanity has recognized the consequences that an impact of an asteroid or comet can have on the development of life on Earth. For this reason, various strategies have been designed to deal with this threat. Among the options analyzed, the deflectinon of the object well before impact is the most promising. One of the methodologies proposed for deflection is a kinetic impact: hitting the body with a massive object to transfer linear momentum that changes its course. The net transfer of momentum depends on many factors, like the size,  internal structure and material properties of the target, as well as the amount and velocity distribution of ejected material. To perform a real test of the kinetic impactor method,in November 2021 NASA launched the DART (Double Asteroid Redirection Test) mission to impact Dimorphos, the satellite of asteroid (65803) Didymos. The impact experiment will occur on 26-Sept 2022 23:14~UT. At  impact time, the DART spacecraft will have a mass of about $550 \ kg$, and a speed of $6.6 \ km/s$ \citep{rivkin2021}. DART will be able to take images during the approach, and send them to Earth, up to a couple seconds before impact. Fifteen days prior to impact, DART will release the  Italian Space Agency's (ASI) ``Cubesat'' LICIACube. LICIACube will follow a similar trajectory, with a closest approach to Dimorphos $167 \ s$ after impact at a distance of $51 \ km$.

The DART impact will likely produce a crater on Dimorphos \citep[or, alternatively, will deform it,][]{raducan2022} and a large amount of material will be ejected at high speed (several tens of $m/s$), producing an ejecta cone that will quickly disperse \citep{fahnestock2022}. The ejecta cone should be observable by LICIACube in the first few seconds after impact. Additionally, during the crater formation phase, low-speed ejecta would be launched from the outer rims of the crater and the surrounding region (up to a few times the impactor size), as predicted from ejecta scaling laws based on the point-source assumption--based scaling models \citep{housen2011}. Moreover, impact generated seismic waves may propagate inside the target and  reach the surface triggering motion of fines that may overcome the escape velocity and produce a late trail of dust \citep{tancredi2012}. The latter effect may take place even in the case of target deformation.

Different sets of data were been used to derive the most relevant physical and orbital parameters for the Didymos-Dimorphos systems. They are summarized in the Didymos Reference Model adopted by ESA. The mean radius of Dimorphos is $r_{Dim}=84 \ m$, and the derived mass is $M_{Dim} = 4.83\times10^9 \ kg$.

Two critical speeds need to be taken into account when discussing the outcome of  low-speed ejecta: \textit{i)} the escape velocity from the surface of Dimorphos: $v_{esc,Dim} \approx 0.089 \ m/s$; and \textit{ii)} the escape velocity from the  binary system (at the distance of Dimorphos from Didymos): $v_{esc,Sys} \approx 0.24 \ m/s$. On one hand, if particles are launched with any outgoing speed $v$ (positive radial component respect to the center of mass of Dimorphos) lower than $v_{esc,Dim}$, they will fall back on Dimorphos. If outwards speed is between $v_{esc,Dim}$ and $v_{esc,Sys}$, particles will be moving in the binary system, and eventually collide with one of the two bodies. Instead, if there is enough time for solar radiation pressure (SRP) to act, small particles can be removed from the system. On the other hand, if a particle's speed is larger than $v_{esc,Sys}$, it will escape from the binary system with a velocity at infinity given ($U$) by: $U^2 = v^2 - v_{esc,Sys}^2$ (without taking into account the effect of SRP). 

Another set of relevant parameters are the acceleration on the surface of Dimorphos due to its self-gravity: $a_{Dim} \approx 4.90\times10^{-5} \ m/s^2$; and the acceleration due to Didymos gravity at the distance of Dimorphos: $a_{Sys} \approx 2.53\times10^{-5} \ m/s^2$.

The goal of this work is to analyze the potential detection conditions of low-speed ejecta from the DART experiment, similarly to the ejection detected in some of the so--called Active Asteroids, or Main-Belt Comets. In Section \ref{lessonsAA}, we discuss some findings regarding this population that are relevant for our study.

To understand  the generation of the low-speed ejecta and its detectability, we  combined the analysis of the following processes, under the assumption of a gravitational aggregate structure for Dimorphos:
\begin{enumerate}
\item Propagation of impact-induced seismic waves at distances far from the impact point.
\item The local effect of these waves into small particles located on the surface, and the ejection mechanism at speeds comparable to the escape velocity of Dimorphos and the Didymos system.
\item Evolution of   ejected particles under the influence of  solar radiation pressure and the gravity field of Didymos and Dimporphos.
\item Prediction of the observation of such cloud of particles from LICIACube and from the Earth.
\end{enumerate}

Hereby, we give detailed information on the studies involved in the processes listed above, the simulations required to get the results and the expected outcome:

\begin{enumerate}
\item
We model the post--impact effects into a gravitational aggregate by means of two independent Discrete Element Method  codes, including   mutual gravity among particles. The hypervelocity impact of the DART projectile is replaced by the impact of a larger, lower mass synthetic projectile, so that conservation of momentum holds and the same amount of residual kinetic energy is delivered to the target. The residual energy, going into the target once the very shattering phase is over, is estimated from laboratory experiments \citep{walker2013}.
We then compute the evolution of particles close to the surface as a function of the angular distance from the impact point (angle Impact-Center-Surface point). 
Section \ref{propagation} is devoted to this item.

\item
We model the incidence of a shock from below into a surface layer of particles. Particles on the upper part of a surface layer will be lofted by a shock at its bottom. We analyze the distribution of velocity of ejected particles and compare it with the escape velocity from the Dimorphos surface ($v_{esc,Dim}$) and from the Didymos  binary system at the distance of Dimorphos ($v_{esc,Sys}$). This is explained in
Section \ref{shaking}.

\item
We analyze the dynamical evolution of particles ejected at slow speed to obtain the velocity distribution when  particles escape the binary system. When  particles enter interplanetary space, they undergo SRP, therefore they are pushed away from the Sun, forming a cometary-like tail, an effect that has already been observed in several so--called Active Asteroids or Main-Belt Comets (see Section \ref{lessonsAA}). 

\item
The outcome from the previous stages can be used to generate synthetic images of the comae and the tails as seen from the Earth at different times after the impact. We also discuss the observation conditions that the LICIACube flyby could encounter.

These two last items were analyzed by \citet{moreno2022} in a companion paper. In Section \ref{observation} we highlight their main conclusions and we include some further analysis of observational consequences of the low-speed ejecta.

\end{enumerate}

In Section \ref{discussion} we discuss the results of the previous items, taking into account the lessons learned from the observations of Active Asteroids.

\subsection{Lessons learned and to be learned from Active Asteroids}\label{lessonsAA}

The small bodies coming to the inner region of the Solar System has been classified in two groups: asteroids and comets. They are observationally distinguished because comets present --at some part of their orbit-- a coma and a tail generated after the release of gas and dust from the solid nucleus; while asteroids are inert. These differences were interpreted as a consequence of material constituents: comets have a mixture of refractory materials and ices, where the latest can sublimate due to the Sun heat; while asteroids are formed with only refractory material, as was confirmed by several space missions to such objects.

This physical distinction has led to identify some properties in the orbital parameters phase-space that can be used to characterize those two populations. A detailed analysis of the orbital classification can be found in \citet{tancredi2014}. Based on this kind of classification, puzzling objects are found: inactive objects with cometary-type orbits (the so-called "Asteroids in Cometary Orbits - ACOs"), and asteroids that showed some kind of cometary-like activity (production of comae and tails), but they are in typical asteroidal orbits (the so-called "Comets in Asteroidal Orbits - CAOs", or "Active Asteroids"). Among the latter group, there are some objects that show recurrent activity events, and they were named as "Main-Belt Comets - MBCs". For a recent review on the topic see \citet{jewitt2022}.

The physical processes associated with the onset of cometary-like activity on Active Asteroids is still a matter of debate. 
We wish to emphasize the "cometary-like activity" term: comet activity is driven by the sublimation of ices and the release of dust embedded in the ice. Instead, it has not been proven so far that  ice sublimation has ever taken place on Active Asteroids (not even on  Main-Belt Comets). In fact, no gaseous species have been detected so far, only the dust in the coma and tail was observed.

Nevertheless, there is consensus among scientists working on this topic, that  activity on several Active Asteroids was generated after impacts and the consequent  release  of large amounts of dust, as was likely the case of asteroids (596) Scheila and P/2016 G1 \citep{jewitt2011,moreno2011,hainaut2019}. From the analysis of the evolution of the tail, it has been concluded that the generation of the long-lasting tails can only be explained if dust is ejected at velocities just above the escape speed of the bodies. Let's take the example of active asteroids P/2019 A4 and P/2021 A5, studied by \citet{moreno2021}. The upper limits for the sizes of these objects are: diameter $D = 0.7-1.0 \ km$ for P/2019 A4, and $D = 2.4 \ km$ for P/2021 A5; although the sizes derived from the dust-formation Monte Carlo modelling are much smaller: $100-240$~m (nominal value $170$ m) for P/2019 A4, and smaller than $500 \ m$ (nominal value $150 \ m$) for P/2021 A5. The estimates of the total dust mass ejected were $(2.0 \pm 0.7) \times 10^6 \ kg$ and $(8 \pm 2) \times 10^6 \ kg$ for P/2019 A4 and P/2021 A5, respectively, for particle size distributions having a maximum particle radius of $1 \ cm$.
The derived ejection speeds for both objects $\sim0.2$~m/s are consistent with the escape speeds at the surface corresponding to  the nuclear radii estimated from the Monte Carlo modelling.

Assuming that the ejected material was produced in a cratering event, \citet{moreno2021} used the cratering scaling--laws laws by \citet{housen2011} with  impact speed of $5 \ km/s$,   ejection speed of $0.23 \ m/s$, and ejected mass of $2\times10^6 \ kg$, to estimate an impactor mass of $200 \ kg$ in the case of P/2019 A4.

In summary, the coma and long-lasting tails observed for a few months on those km-size Active Asteroids are explained by a single impact event of a few hundreds $kg$ projectile, where a few thousands of tons of $mm$ to $cm$-size dust particles are released at speeds just above the escape speed at the surface of those bodies.

Since the comae and the tails of those Active Asteroids were detected long after the generation event, it is not possible to infer the physical mechanism that produced the low-speed ejecta. We also do not know the fraction of high-speed ejecta  produced at impact, because such ejecta was already dispersed at the time of the observation.

The DART experiment can be considered as the generation of an artificial Active Asteroid, with a precise impact date, impact speed and mass of the impactor. Similar observational effects are foreseen: brightness increase of several magnitudes above the nucleus', and production of a long tail. Given that the LICIACube will monitor the impact and the outcome in the few minutes afterwards, and that there will be a large battery of ground-based and space-based telescopes monitoring the event up to several weeks later, we will have the opportunity to get insight on the generation of an Active Asteroid. The DART event  will give useful information for understanding the detailed physics of this phenomena.

\section{Propagation of impact-induced seismic waves} \label{propagation}

\subsection{DART and Dimorphos initial setup}
At this initial stage, we investigate the possible reaction of Dimorphos to the DART collision, under the assumption that it is a spherical gravitational aggregate produced in the formation of the binary system \citep{walsh2008} 
The very internal structure of the target is unknown, therefore, we model the target by a multi--dispersed distribution of $\approx 100\, 000$ spherical particles,  with particle radii ranging from $1.0$ to $2.5 \ m$ and a random packing configuration. The total mass of the particles corresponds to the mass of Dimorphos mentioned above.


We  perform numerical simulations of the DART collision event on Dimorphos using two different Discrete Element Method (DEM) codes (Section \ref{modelling}). We neglect the presence of Didymos, since it does not affect the dynamics on Dimorphos's particles along the small time scale of the collision event, while the momentum and energy propagation takes place. The real DART impact will be a hyper-speed cratering event, where most of the impact kinetic energy goes into the shattering phase, which implies local vaporization, melting, rock deformation, and heat transfer. Therefore, due to the fact that DEM codes are not suitable to simulate the shattering phase itself, we instead concentrate on the effect of the collision on the part of the target not affected by such phase, once it is over. That means that the kinetic energy delivered to the rest of the body is just a small fraction of the impact energy.
Therefore, our synthetic projectile needs to deliver the same nominal linear momentum to Dimorphos as the DART spacecraft will, but it  only delivers to the target a small fraction of the original impact kinetic energy.
According to  cratering experiments \citep{walker2013}, only a small fraction $f_{ke}\approx 0.25 \%$ of impact energy will survive as kinetic energy of the target after the shattering phase.

In order to apply this model, we replace the real DART spacecraft, with  mass $M$ and velocity $V$, with a  synthetic projectile with the same linear momentum as DART and a kinetic energy corresponding to the small fraction surviving the impact. The real DART projectile carries a kinetic energy $E = \frac{1}{2}MV^2 = 1.2\times10^{10} \ J$ and linear momentum $P = M V = 3.6\times10^6 \ kg m/s$. The synthetic projectile has mass $m_0$ and velocity $v_0$. After the impact, we assume that our synthetic Dimorphos will receive a residual kinetic energy of $f_{ke}E$ as a whole, and a linear momentum $P$. The latter includes the linear momentum $P_{ej}$ taken away by the particles ejected and the linear momentum $P_T$ that Dimorphos itself actually received.

However, unlike  reality, our synthetic projectile will not be destroyed by the impact, it will just rebound on the synthetic Dimorphos with a given restitution coefficient $\epsilon$ (which is assumed to be $0.3$, as for typical Earth rocks). Therefore, our synthetic projectile  carries away a kinetic energy $E_{0}=\frac{1}{2}m_{0}(\epsilon v_{0})^2$.
In addition, the initial linear momentum $P_{0}$ of our synthetic projectile must be equal to the linear momentum of Dimorphos itself ($P$) and the ejecta plus the linear momentum taken away by the rebound synthetic projectile ($\epsilon m_{0}v_{0}$). The above description leads to the following equations: 

\begin{equation}
    (1+\epsilon)m_{0}v_{0} = MV
	\label{eq:rescale1}
\end{equation}

\begin{equation}
    \frac{1}{2}m_{0}v_{0}^2 = \frac{1}{2}m_{0}(\epsilon v_{0})^2 + f_{ke}\frac{1}{2}MV^2.
	\label{eq:rescale2}
\end{equation}

By solving such equations, we obtain the following expressions for the mass and velocity of the synthetic projectile, as a function of the parameters $\epsilon$ and $f_{ke}$:

\begin{equation}
    m_{0}=\frac{M}{f_{ke}}\frac{(1-\epsilon)}{(1+\epsilon)}
	\label{eq:result1}
\end{equation}

\begin{equation}
    v_{0}=\frac{f_{ke}V}{(1-\epsilon)}
	\label{eq:result2}
\end{equation}

\begin{table}
    \caption{Caption}
    \label{tab:data}
    \centerline{
    \begin{tabular}{lr} 
       Mean radius of Dimorphos & $84 \ m$ \\
       Total mass of Dimorphos & $4.83\times10^9 \ kg$\\
       Particles sizes & $1-2.5 \ m$ \\
       Particle density & $4,156 \ kg/m^3$ \\
       Projectile radius & $1.485 \ m$ \\
       Projectile mass & $140,000 \ kg$ \\       Projectile density & $10,208 \ kg/m^3$ \\
       Projectile velocity & $23.75 \ m/s$ \\
       Impact point & $20^\circ$ below equator \\
       Projectile velocity vector & pointing radially to the centre
    \end{tabular}}
\end{table}

\subsection{Modelling the outcome of the DART impact \label{modelling}}

To numerically simulate the effects of the DART impact event and the propagation of the induced momentum and kinetic energy waves into the interior of a synthetic Dimorphos, we used two different DEM codes:

\begin{enumerate}
\item The high-performance \textit{PKDGRAV} parallel gravitational N-body numerical code \citep{richardson2000}, with an implementation of the soft-sphere discrete element method \citep[\textit{SSDEM,}][]{schwartz2013} with gravity. \textit{SSDEM--PKDGRAV} has been extensively used in simulations of rubble-pile models of asteroids. The contact model among the particles is a linear spring--dashpot, in which the dashpot force is linearly proportional to the normal and tangential relative velocities. The code also takes into account tangential friction forces  between  contact surfaces. The resistance force acting on the surfaces of the contacting particles will also impose a torque on both particles. By integrating the force and torque, the motion of a particle can then be determined.

\item An extended version of the open-source DEM code \textit{ESyS-Particle} \citep[ https://launchpad.net/esys-particle]{abe2004}, which includes self-gravity between particles. \textit{ESyS-Particle} was first applied in planetary sciences by \citet{tancredi2012}, including simulations in low-gravity environments (asteroids and comets) and new models to simulate contact forces. These early simulations included only a global gravity force in one space direction. In the following articles: \citet{frascarelli2014}, \citet{rocchetti2018}, \citet{nesmachnow2019} and \citet{rocchetti2021}, the authors described a self-gravity module that applied High Performance Computing techniques to enable simulations of hundred of thousands particles efficiently. Different strategies to compute long-range forces were introduced, implemented, and evaluated in realistic scenarios. This extended version of \textit{ESyS-Particle} which include self-gravitation is called \textit{ESys-Gravity}. In \textit{ESyS} the user can select the contact model among different alternatives. In our simulations we have used a linear spring-dashpot and a Hertzian viscoelastic interaction model with friction.

\end{enumerate}

We performed numerical simulations using the same initial conditions for both codes. For the nominal configuration we assumed an efficiency factor $f_{ke}=0.0025 \ (0.25\%)$ and a restitution coefficient $\epsilon = 0.3$. Using \ref{eq:result1} and \ref{eq:result2}, we computed the mass and velocity of the projectile, which are listed in Table \ref{tab:data}. The impact occurs at a point on the surface $20 ^\circ$ below the equator, and the projectile velocity vector points radially towards the centre of the target.

Input parameters related to elastic properties are different among the two codes. \textit{PKDGRAV} uses the restitution coefficient ($\epsilon~=~0.3$) and a given maximum allowed overlap during a collision ($\delta R = 3\%$), as well as suitable time step and elastic constant, dependent on the system characteristics. Namely, during the first 3 seconds of the impact event, due to the high speed collision from the synthetic DART impactor, we use a short time step of $3\times10^{-5}$ s  and a spring constant of $1.58\times10^{12} \ kg s^{-2}$. Between 3 seconds and 18 minutes, we switch to a time step of $0.01$ s and a spring constant of $2.13\times10^{6} \ kg s^{-2}$ since collisions in that time range period are less energetic. Finally, for the rest of the simulation, we adopt a time step of $0.055$ s and spring constant of $8.67\times10^{4} \ kg s^{-2}$.
For the nominal case, in \textit{ESyS-Gravity} we use the linear spring-dashpot model with a spring constant of $10^9 \ Pa$ and damping coefficient $10^5 \ (m s)^{-1}$; which produces a restitution coefficient similar to the one assumed by PKDGRAV in experiments like the bouncing of two $1 \ m$ equal spheres.

\begin{figure}
    \centering
    \includegraphics[width=0.45\linewidth]{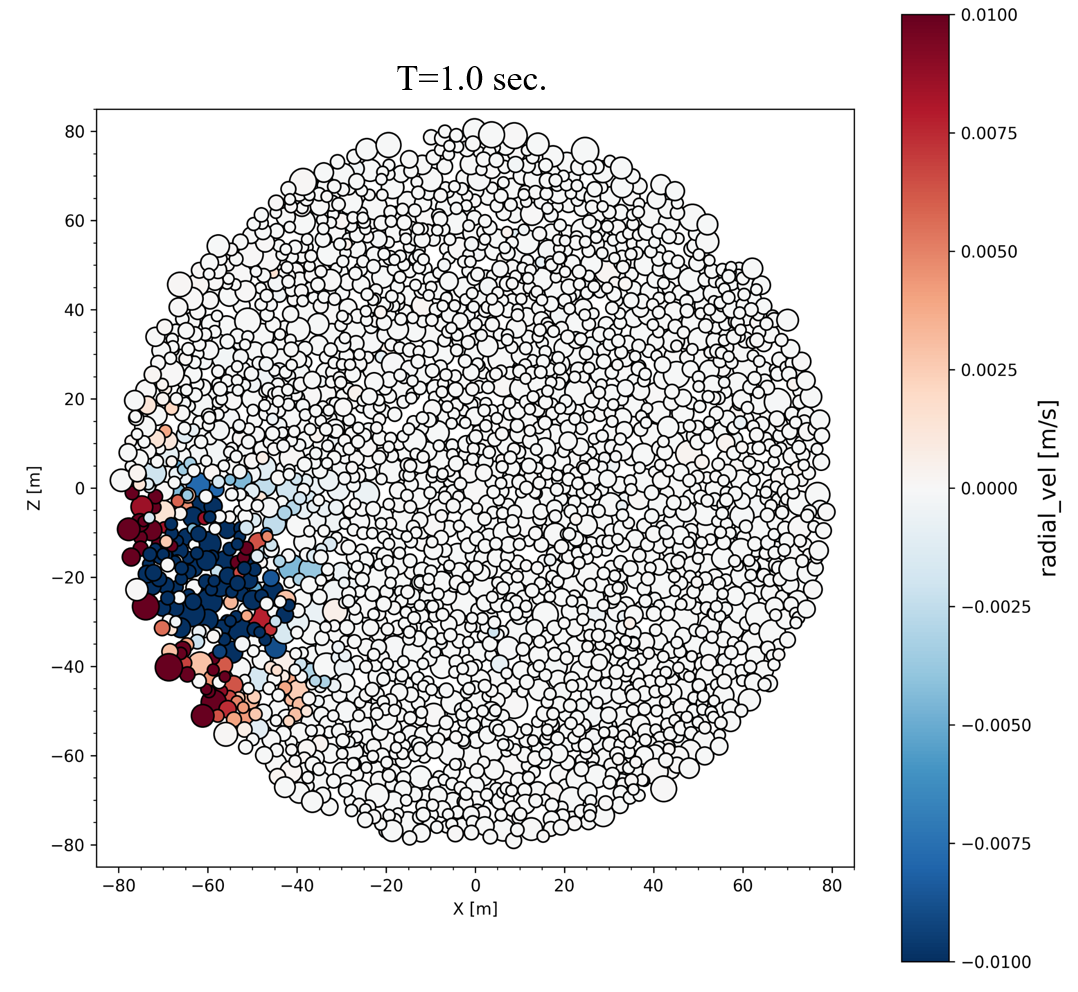}
    \includegraphics[width=0.45\linewidth]{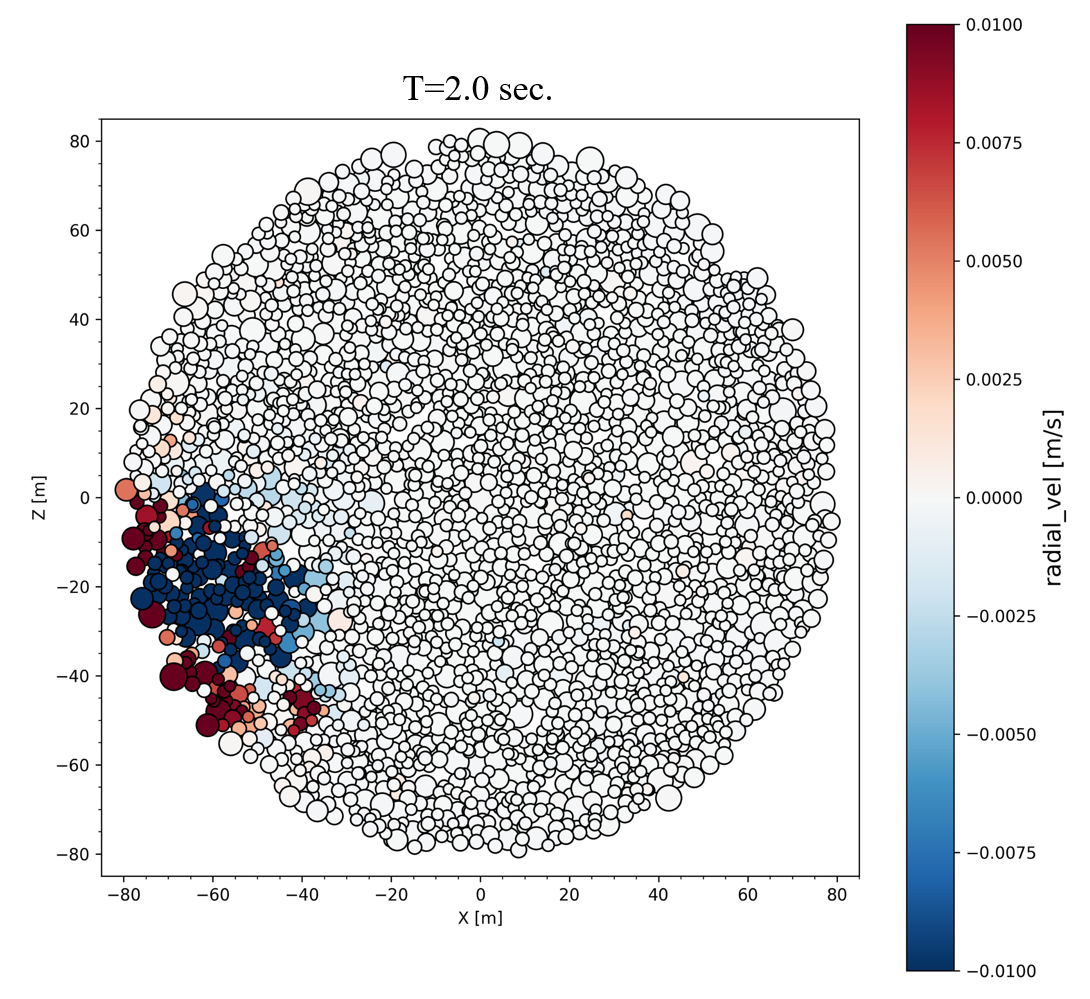}
    \includegraphics[width=0.45\linewidth]{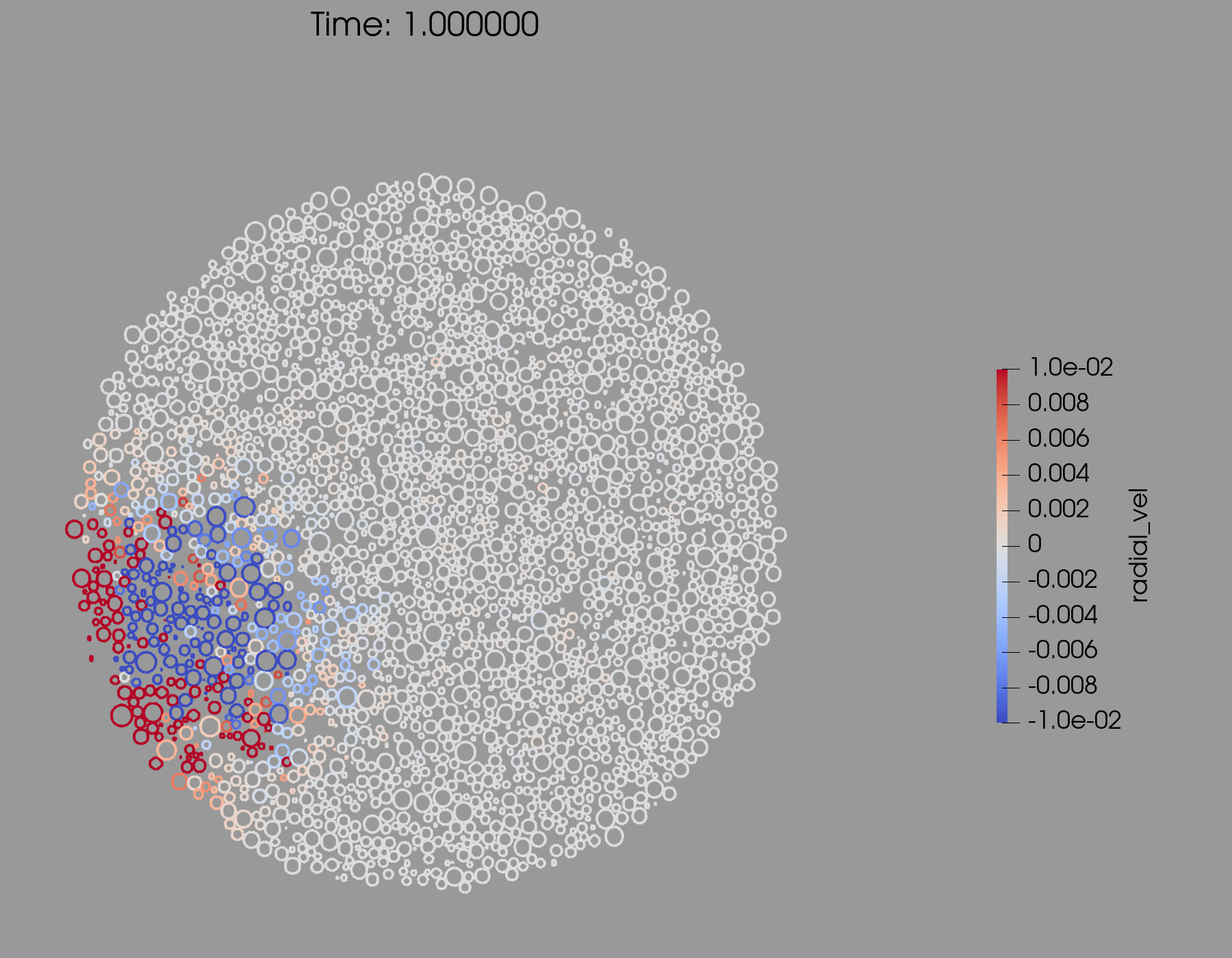}
    \includegraphics[width=0.45\linewidth]{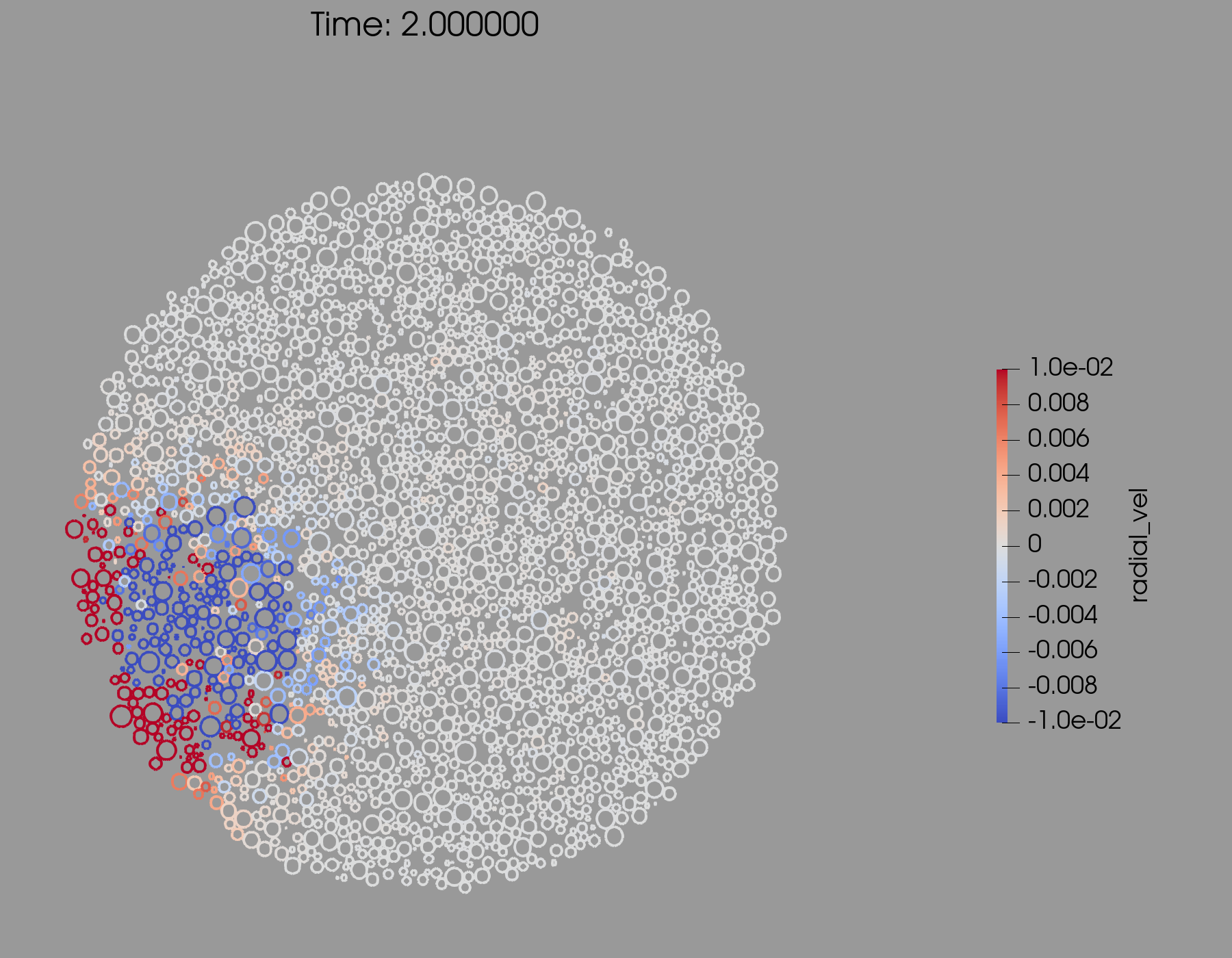}
    \caption{Snapshots of a cross section through the center of the target and the impact point. The upper row corresponds to simulations with \textit{SSDEM--PKDGRAV}, and the lower row to simulations with \textit{ESyS-Gravity}. The left-hand plots are snapshots at $1 \ s$ after impact, and the right-hand plots at $2 \ s$. The color bar is proportional to the value of $v_{rad}$, the projection of the particle's velocity vector along the radius vector of the particle. Red color corresponds to outward velocities (positive values), and blue color to inward velocities (negative values).}
    \label{fig:slice}
\end{figure}

\begin{figure}
    \centering
    \includegraphics[width=0.45\linewidth]{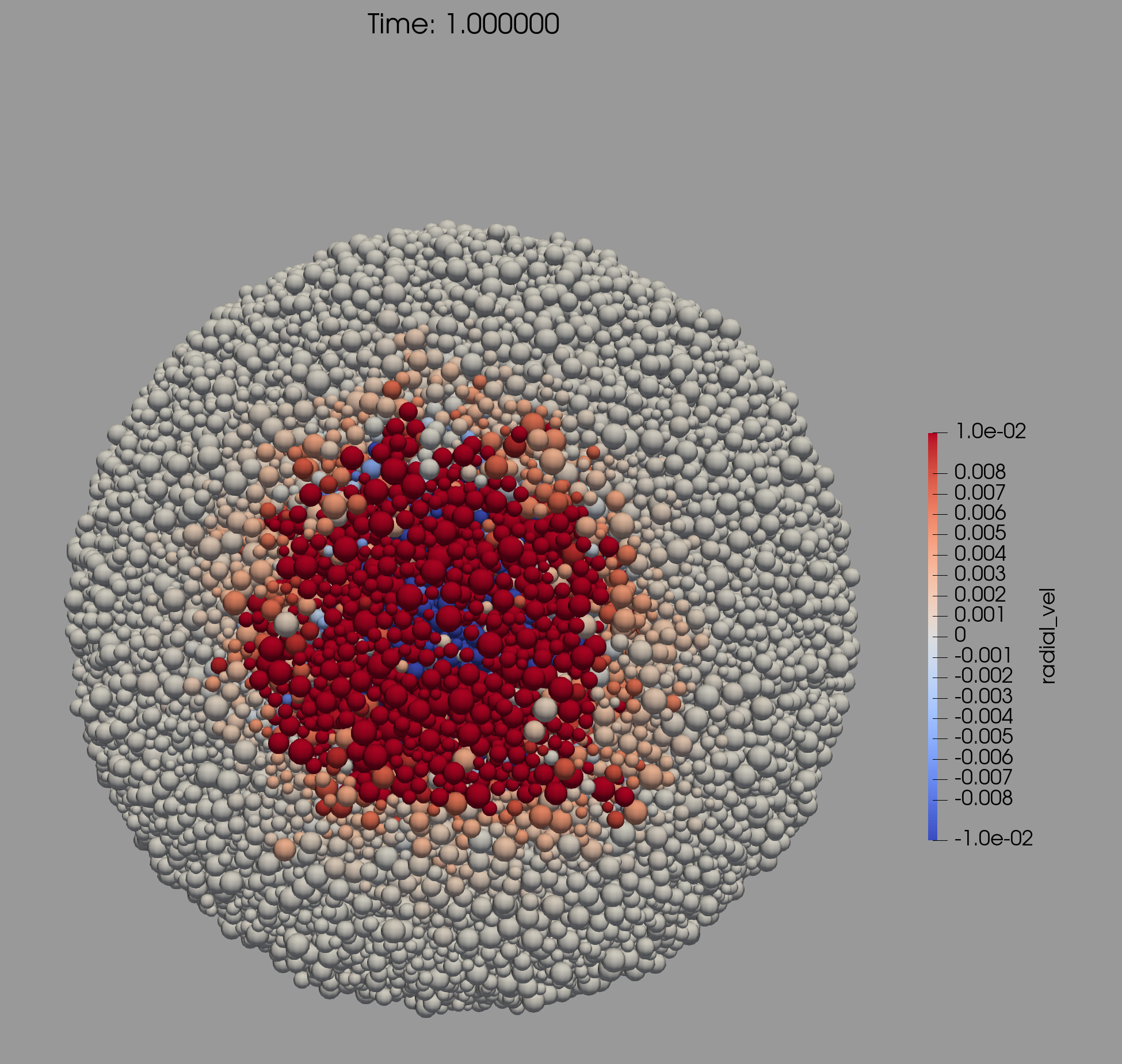}
    \includegraphics[width=0.45\linewidth]{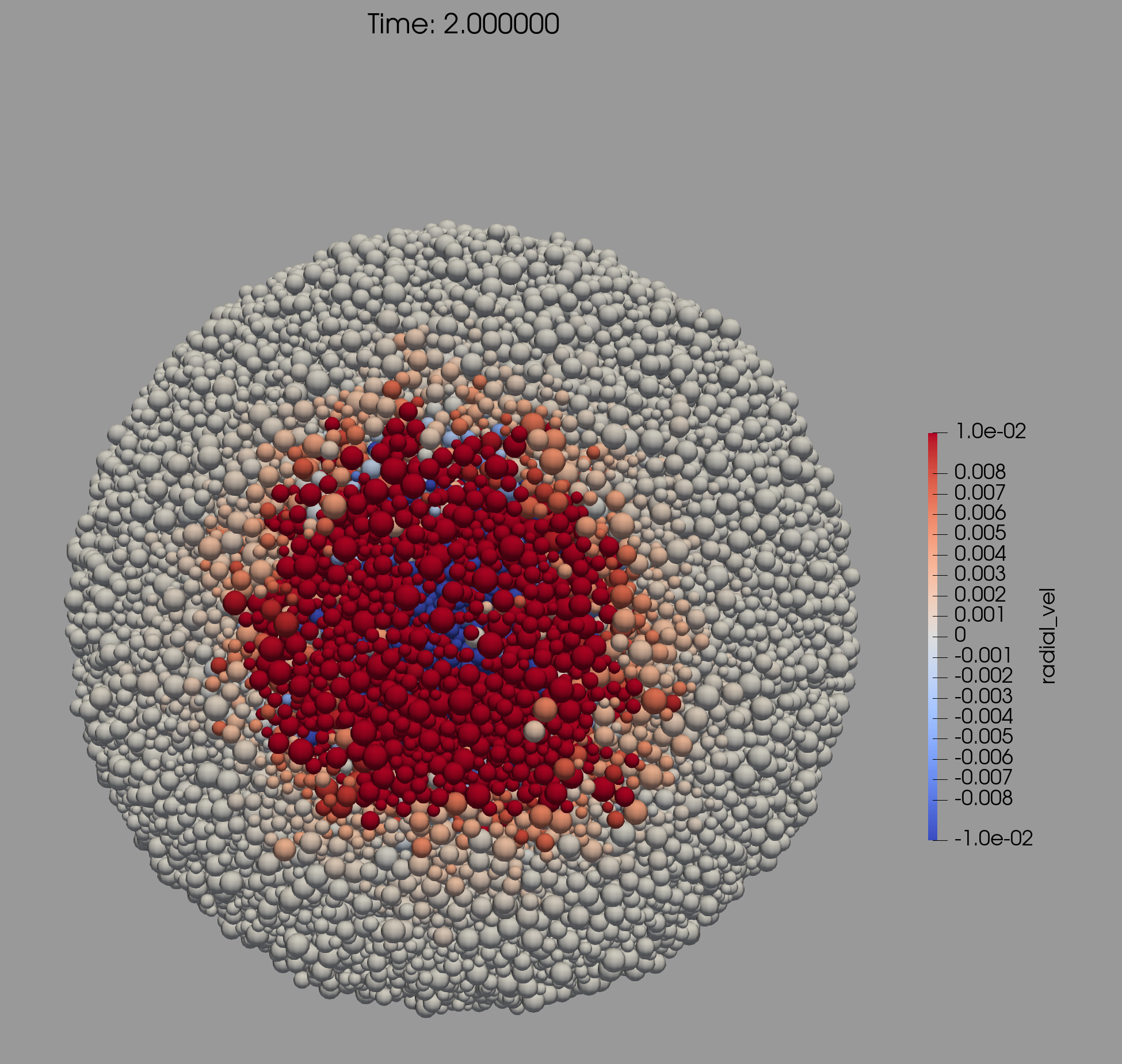}
    \caption{Snapshots at 1 and 2 seconds after the impact, from a viewpoint located above the impact point. The simulations were done with \textit{ESyS-Gravity}. The color bars are the same as the ones used in Fig. \ref{fig:slice}.}
    \label{fig:topview}
\end{figure}

\subsection{Summary of results of energy and momentum propagation} \label{resprop}

As a consequence of the propagation of the impact--induced seismic wave, there is a large scatter between the evolution of the particles on the surface and the interior.

In Fig. \ref{fig:slice} we present snapshots at 1 and 2 seconds after the impact of a cross section through the center of the target and the impact point. The color bar is proportional to the value of $v_{rad}$, the projection of the particle's velocity vector along the radius vector of the particle. Red color corresponds to outward velocities (positive values), and blue color to inward velocities (negative values). The upper plots correspond to simulations using \mbox{\textit{SSDEM--PKDGRAV}}, while the lower ones to simulations using \textit{ESyS-Gravity}.

We do not go into a detailed comparison of both sets of simulations, but we observe that the velocity field pattern in the interior looks very similar in both models.

Let's recall that the impact point in both models is 20$^\circ$ below the equatorial plane (lower left part in each snapshot of Fig. \ref{fig:slice}). 2~seconds after impact (upper and lower right--hand plots),  particles moving inward (blue color scale)  reached up to about half the radius of the target. We observe that the seismic wave going to the interior bounces back, and produces outward displacement on the surface, and inward displacement in the interior. The range of speed values is the same in both models. The volume affected by the impact induced wave is a half sphere, centered on the impact point, with a radius approximately half the body radius. Note that  particles on the surface are moving outwards (red colors).  

We conclude that the results obtained with the two models are comparable. Hereafter, we will present the results using only \textit{ESyS-Gravity}, as it can be later applied to the propagation of linear momentum and kinetic energy all the way to small particles on the surface (Section \ref{shaking}).

We are mainly interested in the effects on the surface particles at locations several times away from the expected crater radius.

To analyze the effects on the particles far from the impact point, we consider 3 vector parameters: the displacement ($\Delta P$) with respect to the original position ($P$), the velocity ($v$), and the acceleration ($a$). We compute the radial projection of these vector parameters, \textit{i.e.} the dot product of $\Delta P_{rad} = \Delta P \cdot P$, $v_{rad} = v \cdot P$ and $a_{rad} = a \cdot P$.

In Fig. \ref{fig:topview} we present snapshots at 1 and 2 seconds after the impact, from a viewpoint located above the impact point. The color bars are the same as the ones used in Fig. \ref{fig:slice}. Note that at distances of 30-40$^\circ$ from the impact point, particles are moving outward with speeds over $10^{-2} \ m/s$.

\begin{figure}
    \centering
    \includegraphics[width=1\linewidth]{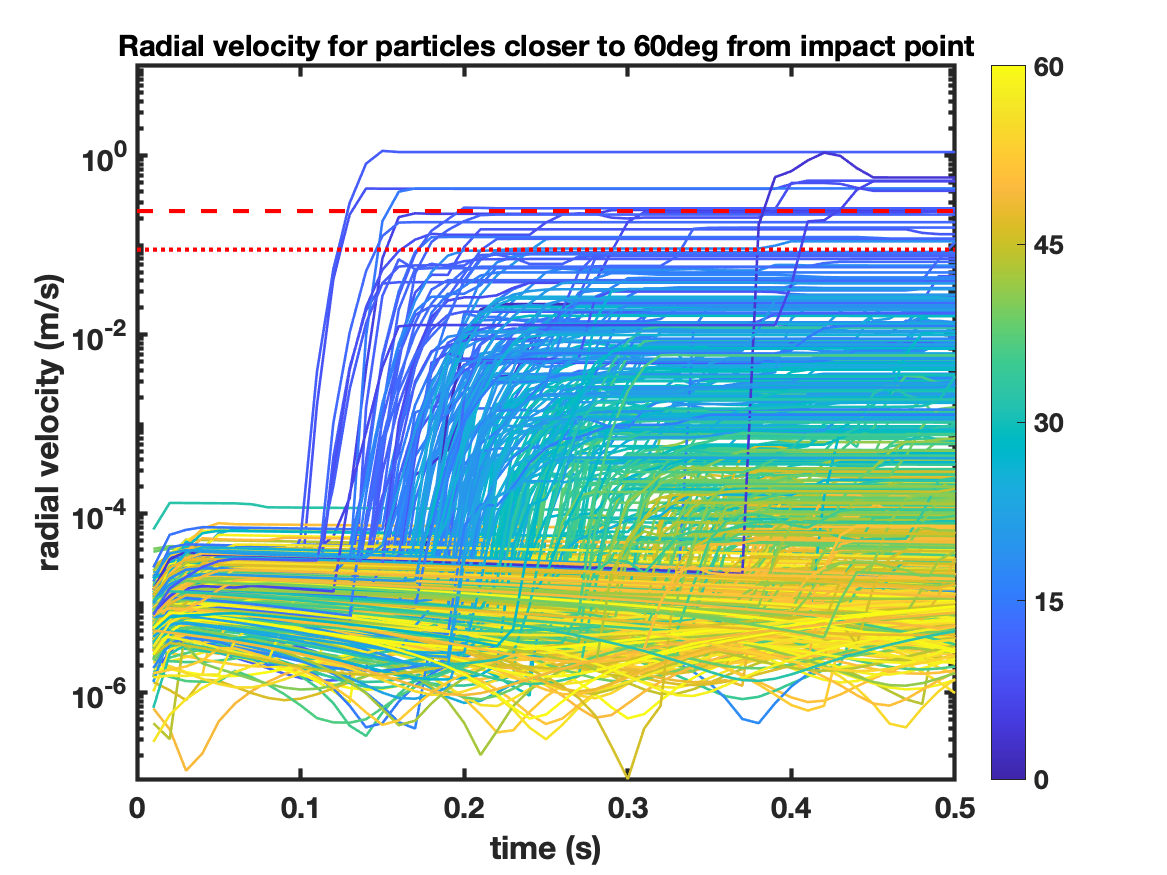}
    \caption{The evolution of the speed for over 600 particles closer than 60$^\circ$ from the impact point and at depth up to $7 \ m$ below the surface.}
    \label{fig:vel}
\end{figure}

\begin{figure}
    \centering
    \includegraphics[width=0.9\linewidth]{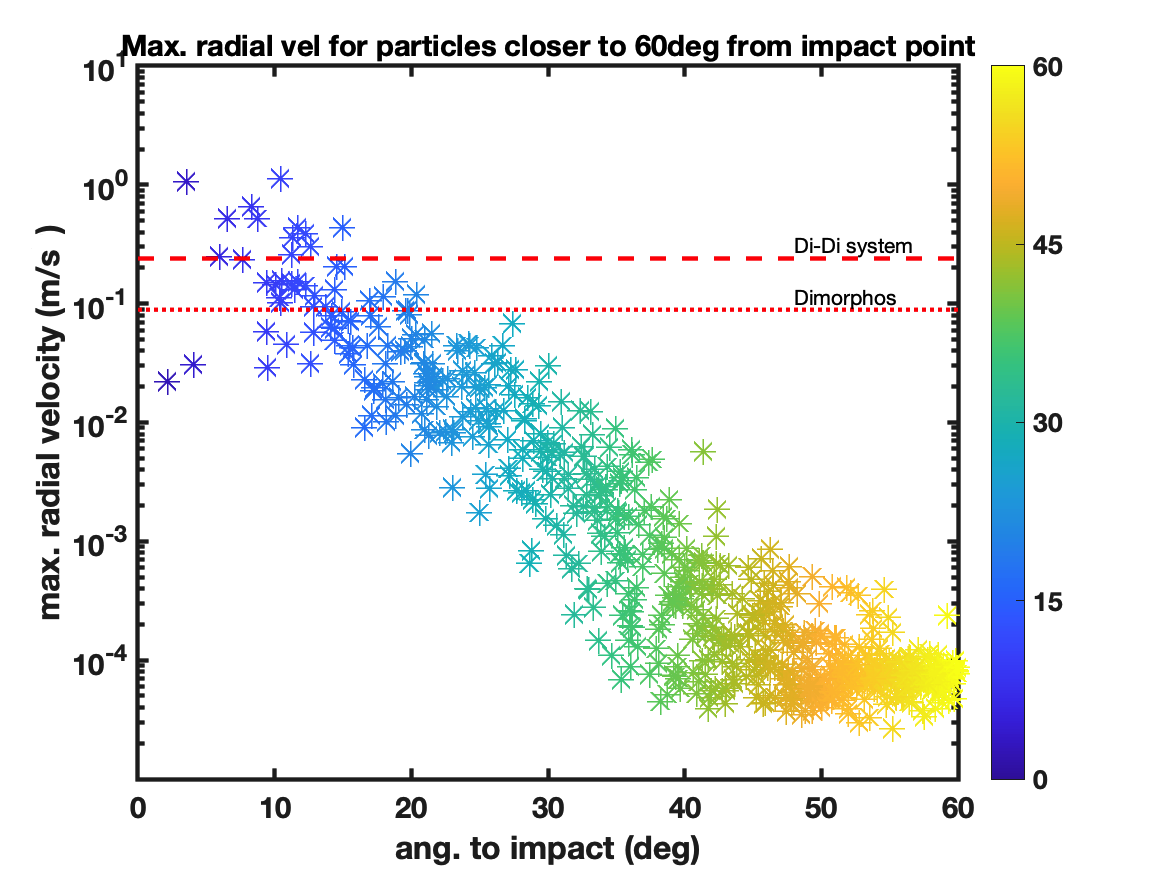}
    \includegraphics[width=0.9\linewidth]{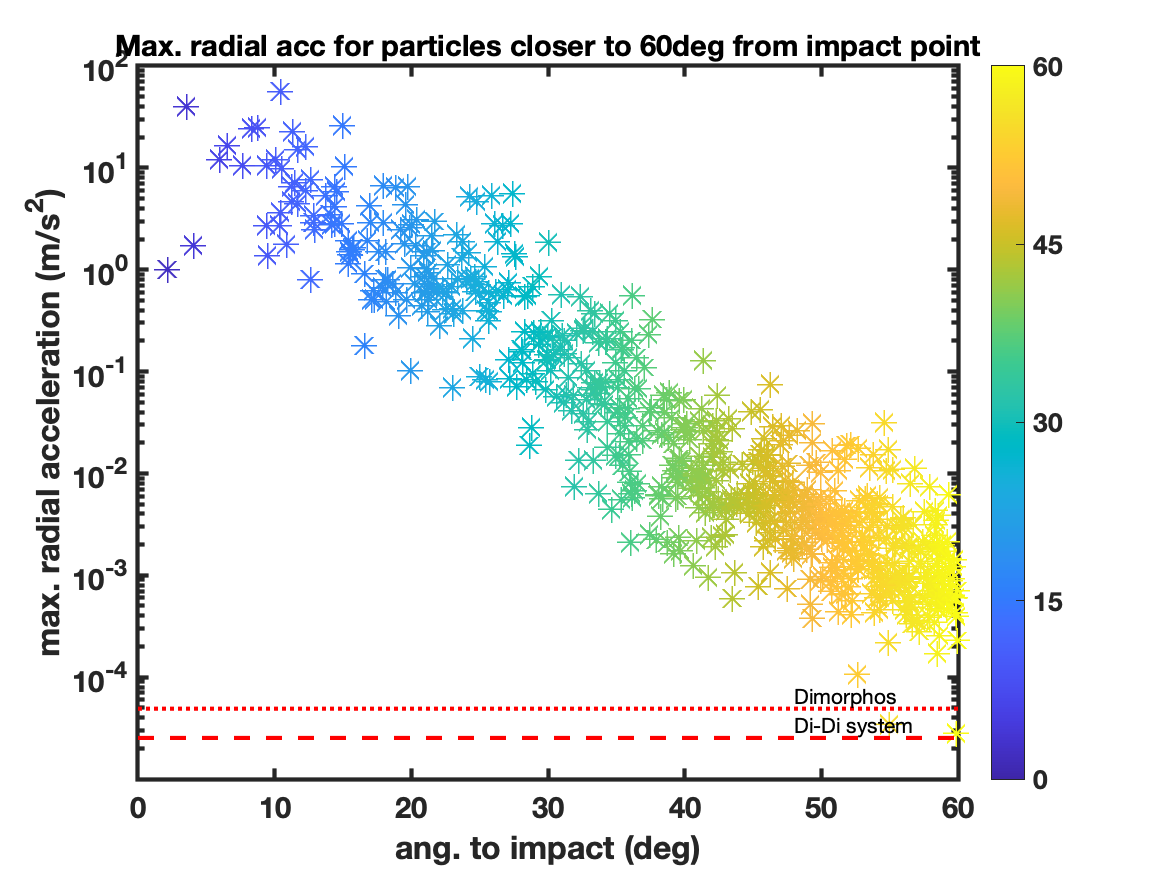}
    \caption{\textit{a)} The maximum outward radial velocity as a function of the surface angle to the impact point (up to 60$^\circ$). The red dotted line corresponds to the escape speed at the surface of Dimorphos ($v_{esc,Dim}$); and the red dashed line to the escape speed from the  binary system at the distance of Dimorphos from Didymos ($v_{esc,Sys}$).
    \textit{b)} The  maximum outward radial acceleration as a function of the surface angle to the impact point (up to 60$^\circ$). The red dotted line corresponds to the gravity acceleration on the surface of Dimorphos ($a_{Dim}$); and the red dashed line, to the acceleration due to gravity by Didymos at the distance from Dimorphos ($a_{Sys}$).}
    \label{fig:angle}
\end{figure}

\begin{figure}
    \centering
    \includegraphics[width=1\linewidth]{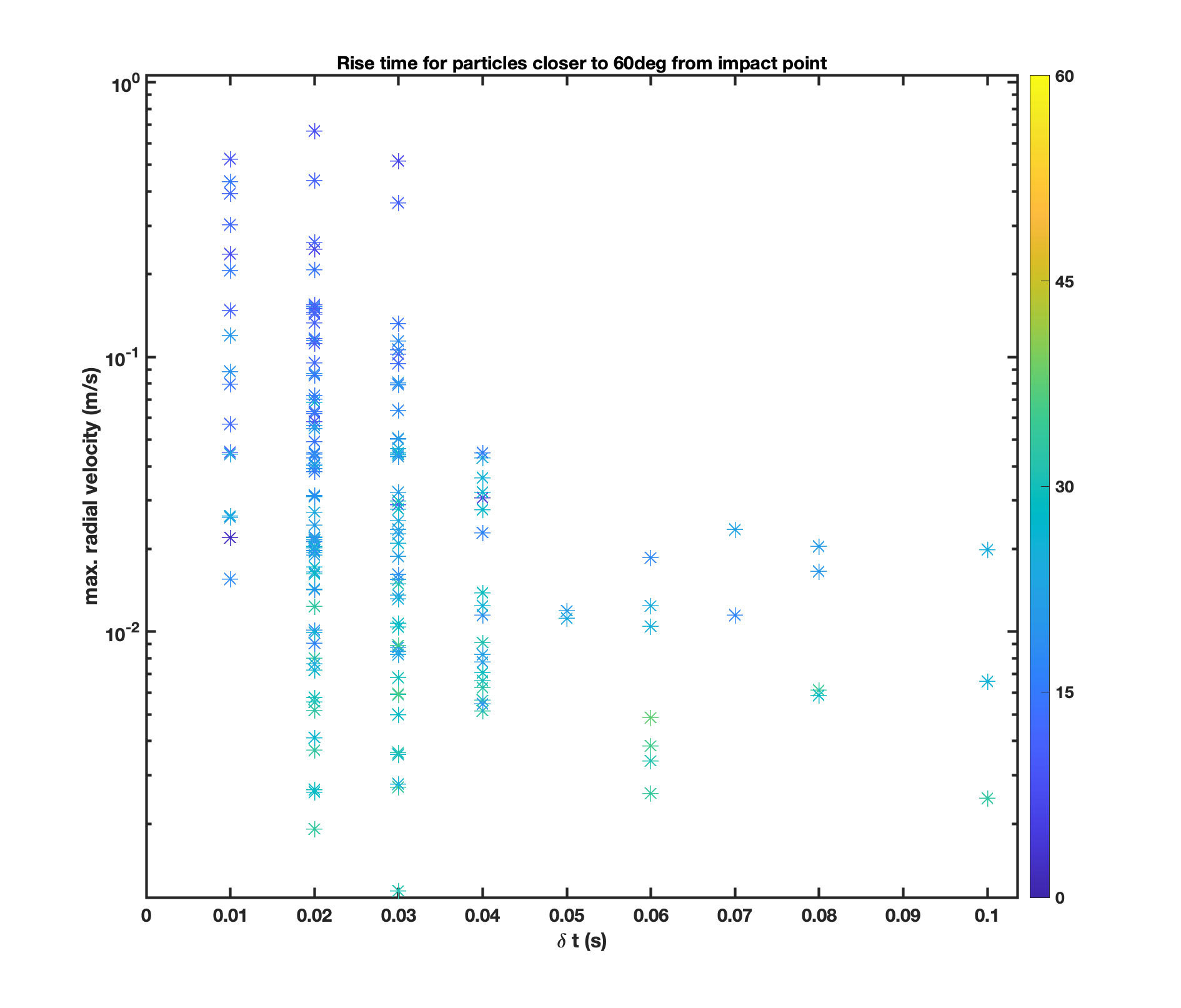}
    \caption{The rise time as a function of the maximum radial velocity. The rise time of the velocity is computed as the time it takes to go from $v < 10^{-4} \ m/s$ to $v > 10^{-3} \ m/s$.}
    \label{fig:rise}
\end{figure}

We focus our analysis on particles that are close to the surface. Particles that are up to $7\ m$ below the surface are selected. The evolution of the speed for over 600 particles closer than 60$^\circ$ from the impact point is shown in Fig. \ref{fig:vel}. In spite of the messiness of this plot, we note that the increase in speed is a sharp step-wise function. In  Fig. \ref{fig:angle}\textit{a} we plot the maximum outward radial velocity, and in Fig. \ref{fig:angle}\textit{b} the  maximum outward radial acceleration, as a function of the surface angle to to the impact point (up to 60$^\circ$). In Fig. \ref{fig:angle}\textit{a} the red dotted line corresponds to the escape speed at the surface of Dimorphos ($v_{esc,Dim}$); and the red dashed line to the escape speed from the  binary system at the distance of Dimorphos from Didymos ($v_{esc,Sys}$). In Fig. \ref{fig:angle}\textit{b} the red dotted line corresponds to the gravity acceleration on the surface of Dimorphos ($a_{Dim}$); and the red dashed line, to the acceleration due to gravity by Didymos at the distance from Dimorphos ($a_{Sys}$).
The rise time of the velocity is computed as the time it takes to go from $v < 10^{-4} \ m/s$ to $v > 10^{-3} \ m/s$. The rise time as a function of the maximum radial velocity is shown in Fig. \ref{fig:rise}.

We conclude that for particles up to 30$^\circ$ from the impact point, a velocity of $\sim 10^{-2} \ m/s$ is acquired in an interval of $\sim 0.02$ s. This corresponds to an acceleration of $a = \delta v / \delta t \sim 0.5 \ m/s^2$; which is related to the acceleration behavior shown in Fig. \ref{fig:angle}\textit{b} for particles at that distance. 

\begin{figure}
    \centering
    \includegraphics[width=1\linewidth]{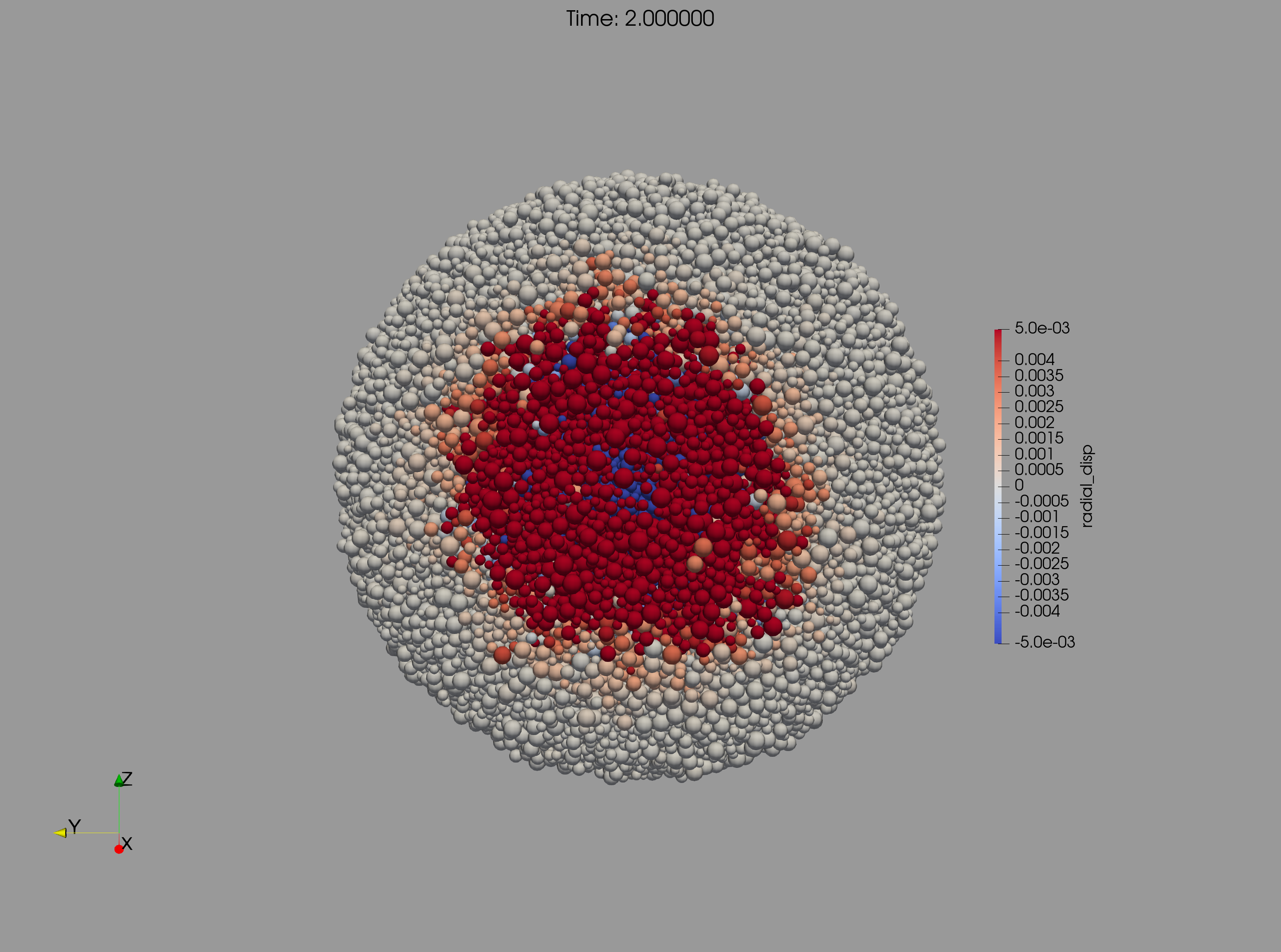}
    \caption{Radial displacement of surface particles. The color bar is proportional to the value of the radial displacement, the projection of the displacement vector along the radius vector of the particle. Red color corresponds to outward displacements (positive values), and blue color to inward displacements (negative values).}
    \label{fig:disp}
\end{figure}

Note that peak outward velocities up to $10^{-3} \ m/s$ are reached by particles up to 40$^\circ$ from the impact point. Also, peak outward accelerations well above the surface acceleration on Dimorphos are attained at distances over 60$^\circ$ from the impact point.

The propagation of the impact induced waves produces radial displacements of $1 \ m$ size particles by over $1 \ mm$ at distances of $40-60^\circ$ from the impact point (Fig \ref{fig:disp}).

We observed that a spherical cap with an angle up to 60$^\circ$ from the impact point is affected by the propagation of the impact induced waves. This caps corresponds to an area $\sim 1/4$ of the total surface area of Dimorphos (\textit{i.e.} $\sim 2\times10^4 \ m^2$).

Finally, we want to note that we assumed a conservative estimate of the efficient coefficient for the transmission of kinetic energy from the impactor to the particles ($f_{ke}=0.25\%$); however, there is a big uncertainty in this value. If the efficient coefficient is larger, the affected area should increase. Other free parameters that could affect the results are the assumed elastic constants and the impact geometry.

Therefore, we cannot rule out that the shaking mechanism could affect an even larger area. It could reach the hemisphere opposite to the impact point, all the way to the antipodal point.

\section{Lofting of particles due to shaking}\label{shaking}

\subsection{The lofting mechanism}

In this section we analyze the arrival of the impact-induced seismic wave  at points located far from the impact point, which produces a shaking effect on the ground. A preliminary analysis of this effect was done in \citet{tancredi2012}.

In  the simulations described in the previous section, we considered $m$-size particles. We observed that $mm$ outward displacements of surface particles in a time scale of hundredth of a second can be produced by the impact-induced wave. Let's consider a regolith layer of small $mm$ to $cm$--size particles on top of the large $m$--size ones. The shaking produced on the floor of the regolith layer will propagate up to the surface, and it will loft the small particles on the upper part of the layer. This is a well-known effect in common--day life. Take a can of fine powder like chocolate powder (cocoa) or flour. Strongly tap it from the bottom of the can, and you will see powder that is lofted from the surface. Due to the strong Earth's gravity, the dust particles immediately fall back. We call this the "cocoa effect" (realized by G.T. when he was preparing breakfast for his kids).

\begin{figure*}
    \centering
    \includegraphics[width=\textwidth]{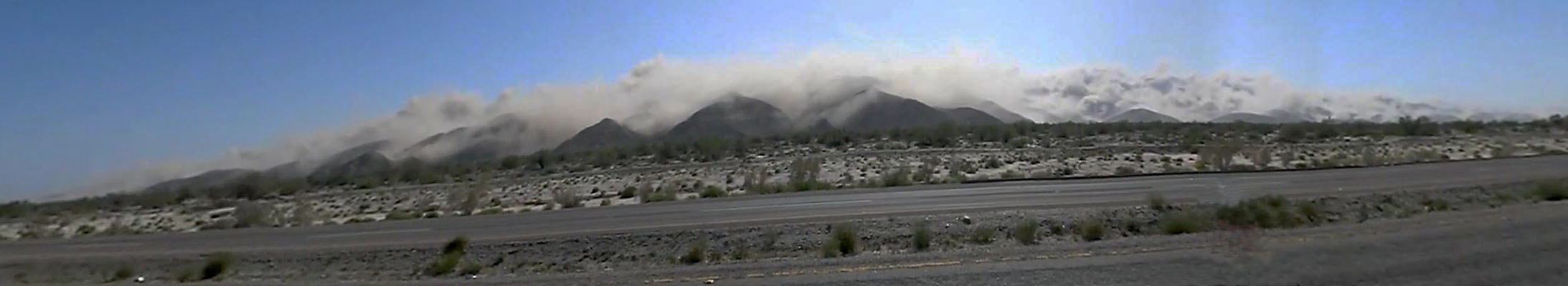}
    \caption{Thick dust clouds were formed over the peaks of Sierra El Mayor immediately after the magnitude $7.2 \ Mw$ earthquake occurred in 4 April 2010 in Northern Baja California (Mexico).The composite photo was created from snapshots of a video taken by Sarah Klaus at $30 \ km$ south of Mexicali, on the main highway from Mexicali to San Felipe.}
    \label{fig:mexicali}
\end{figure*}

This effect can also be observed in a more natural and geophysical environment. An earthquake produces strong shaking on the ground. If an earthquake happens in a desert and dusty area, a great amount of lofted dust is immediately produced after the arrival of the seismic waves due to  shaking and land-sliding. Such effect was observed  in the Mexicali earthquake in April 4$^{th}$, 2010 \citep[][and Fig. \ref{fig:mexicali}]{tancredi2010}. Due to the small size of dust particles and the suspension in the atmosphere, the cloud persisted for several minutes, until it was dispersed by the wind. The propagation of the dust cloud was detected in satellite images.

The physics behind the "cocoa effect" is out of the scope of this paper. We have been performing laboratory experiments as well as numerical simulations to understand it, and to find the relevant laws; \textit{e.g.} the amount and velocity of   lofted dust particles as a function of the floor's displacement. These results will be the matter of a future publication. In this paper, we just present some numerical simulations relevant to our problem.

Note that this mechanism is different from the crater and plume formation. In the formation of the crater, a volume many times the size  of the projectile is excavated and a fraction of it is ejected by forming a plume with a conical shape. The conical plume has a typical half-angle of 45$^\circ$. The physics of the crater and the plume are represented by the point source impact scaling laws by \citet{housen2011}. From these laws, one can obtain the size of the crater, the amount of ejected material and its velocity distribution, depending on parameters like the projectile mass and speed and the mechanical and gravity properties of the target.

The effect that we are discussing here, although related, is  a different mechanism. The area affected by the impact can no longer be considered as a point source, since it is many times larger than the crater. The lofted material is not confined in a conical plume, as it comes from a large area.

In order to analyze the relevance of the "cocoa effect" for such problem, we performed some numerical simulations of the local effect on the surface using ESyS-Particle (it is not necessary to use ESyS-Gravity, because we do not need  self-gravity among   particles; a constant value for vertical gravity acceleration is enough).

\begin{figure}
    \centering
    \includegraphics[width=.8\linewidth]{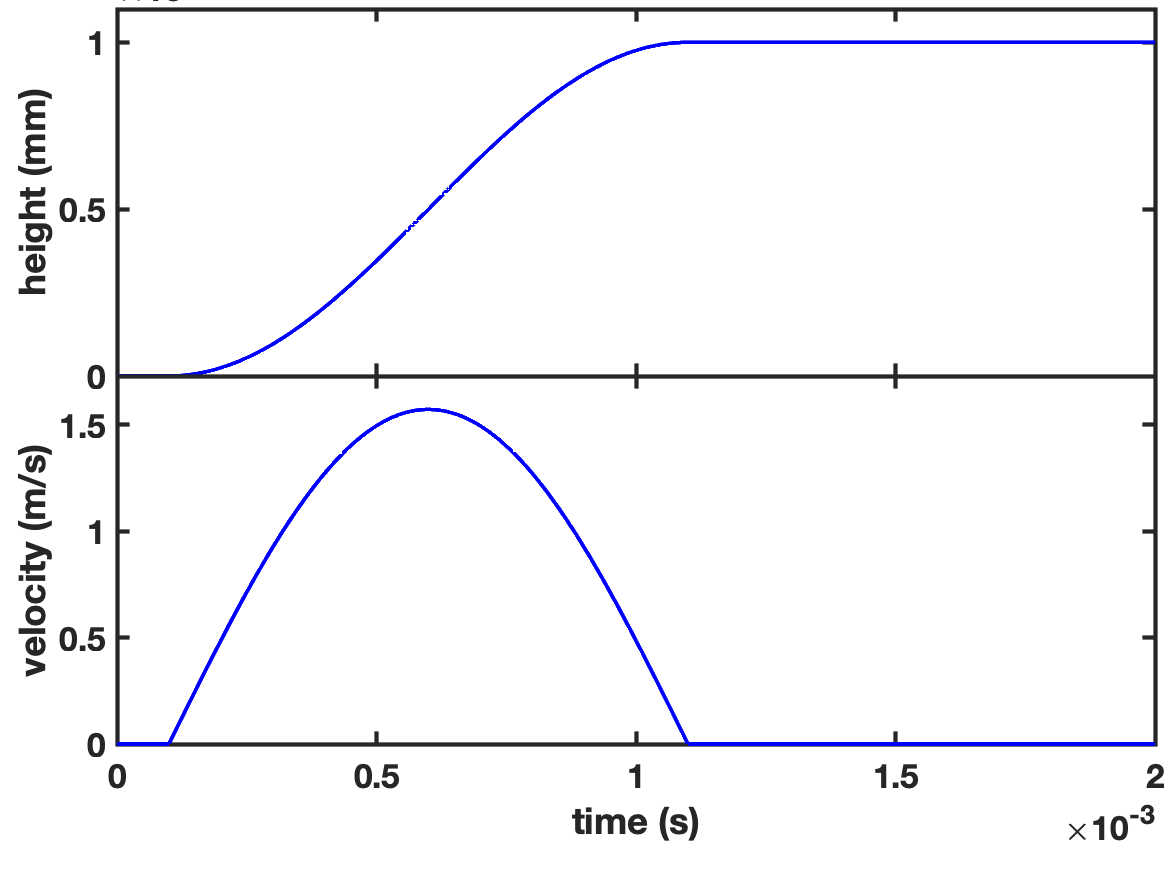}
    \caption{The displacement following a $\sin^2$ function, with variable amplitude and duration. The height of the floor $h$ as a function of time $t$ follows the equation: $h = A \sin^2 \left( \frac{\pi}{2 P} \left (t - dt_{0}) \right) \right)$, where $A$ is the maximum amplitude of the displacement ($A = 1 \ mm$), $P$ is the duration ($P = 10^{-3} \ s$), and $dt_{0}$ is the initial time separation ($dt_{0} = 10^{-4} \ s$). 
    \textit{a)} The height of the floor as a function of time. \textit{b)} The velocity respect to the floor as a function of time}
    \label{fig:floordisp}
\end{figure}

Hereby, we describe the experimental set-up. Consider a cubic box ($4 \ cm$ size) filled with $\sim 300,000$ particles, with radius between $0.2$ and $0.5 \ mm$, and material density of $3000 \ kg/m^3$. One layer of particles was glued to the floor. The interaction among particles is visco--elastic Herztian with friction (see \citet{tancredi2012} for the description of the model and parameters: Young modulus $Y=10^{10} \ Pa$, Poisson ratio $\nu=0.3$, dissipative constant $A=2e-7 \ s^{-1}$, dynamic friction parameter $\mu=0.6$). The interaction of particles with the walls is elastic ($Y=5^{9} \ Pa$). The gravity acceleration is in the vertical direction, with $g= 10^{-3} \ m/s^2$. A little after the beginning of the simulation, we apply to the floor a displacement following a $\sin^2$ function, with variable amplitude and duration, like the one shown in Fig. \ref{fig:floordisp}. The characteristic of the experiments are summarized below:

\begin{itemize}
    \item A cubic box ($4$ cm size) filled with $\sim 300,000$ particles
    \item Particles radius: $0.2-0.5$ mm - Density $= 3000 \ kg/m^3$
    \item One layer of particles was glued to the floor.
    \item Interaction among particles: Viscoelastic Herztian with Friction ($Y = 10^{10} \ Pa$). Elastic Interaction with walls.
    \item Gravity: $g= 10^{-3} \ m/s^2$
    \item A displacement of the floor following $\sin^2$ function, with variable amplitude.
\end{itemize}

\begin{figure}
    \centering
    \includegraphics[width=1\linewidth]{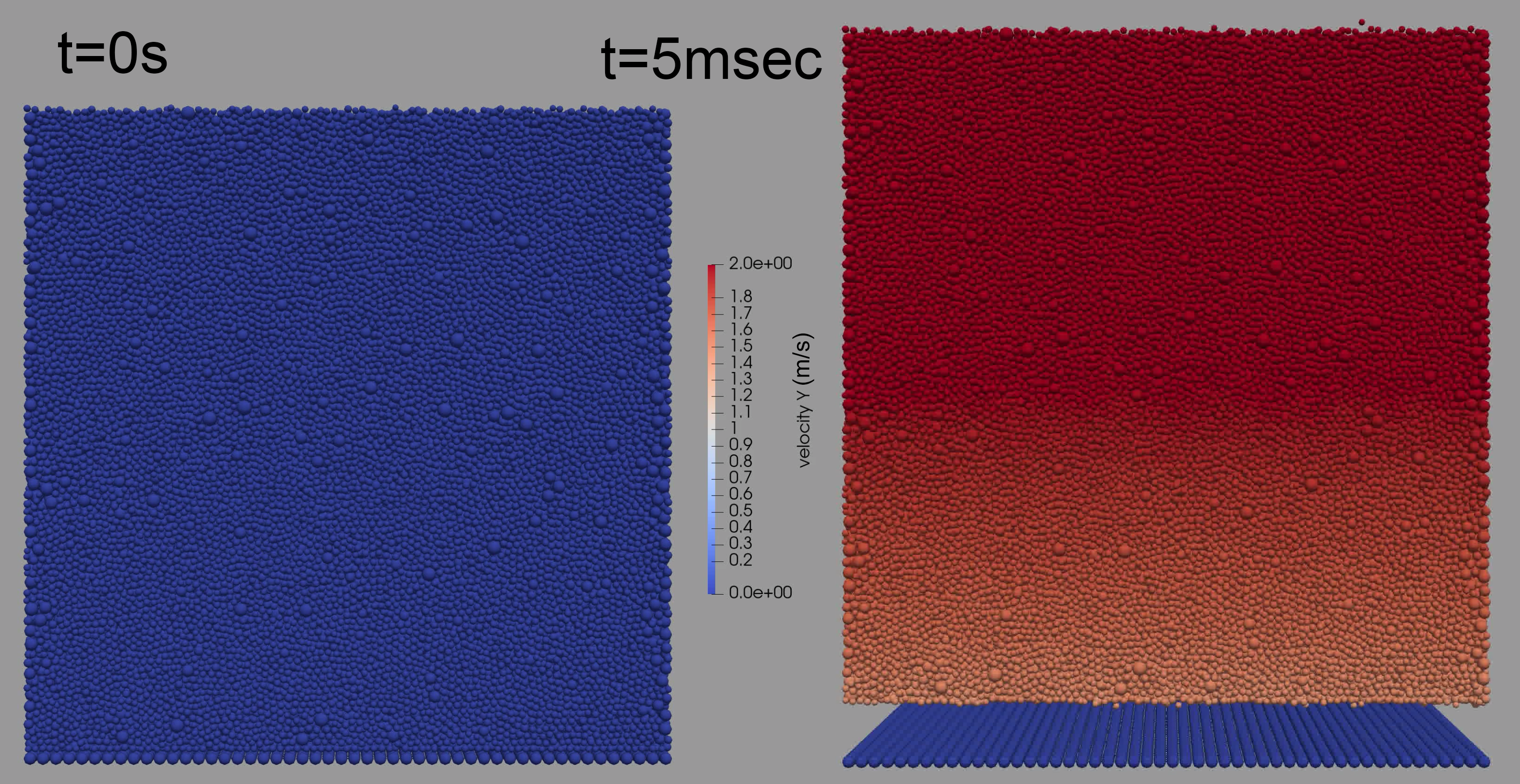}
    \caption{\textit{a)} The initial status of the box with  particles at rest. After we apply the $\sin^2$ displacement to the floor, the particles glued with the floor move upwards, and they knock the particles above. This triggers  particle motion upwards, but due to mutual collisions, there are  differential velocity displacements. The particles on the upper layer move  faster than the lower ones, as  can be seen in Fig. \textit{b)} at 5 milliseconds after the start.}
    \label{fig:shaking}
\end{figure}

\begin{figure}
    \centering
    \includegraphics[width=1.\linewidth]{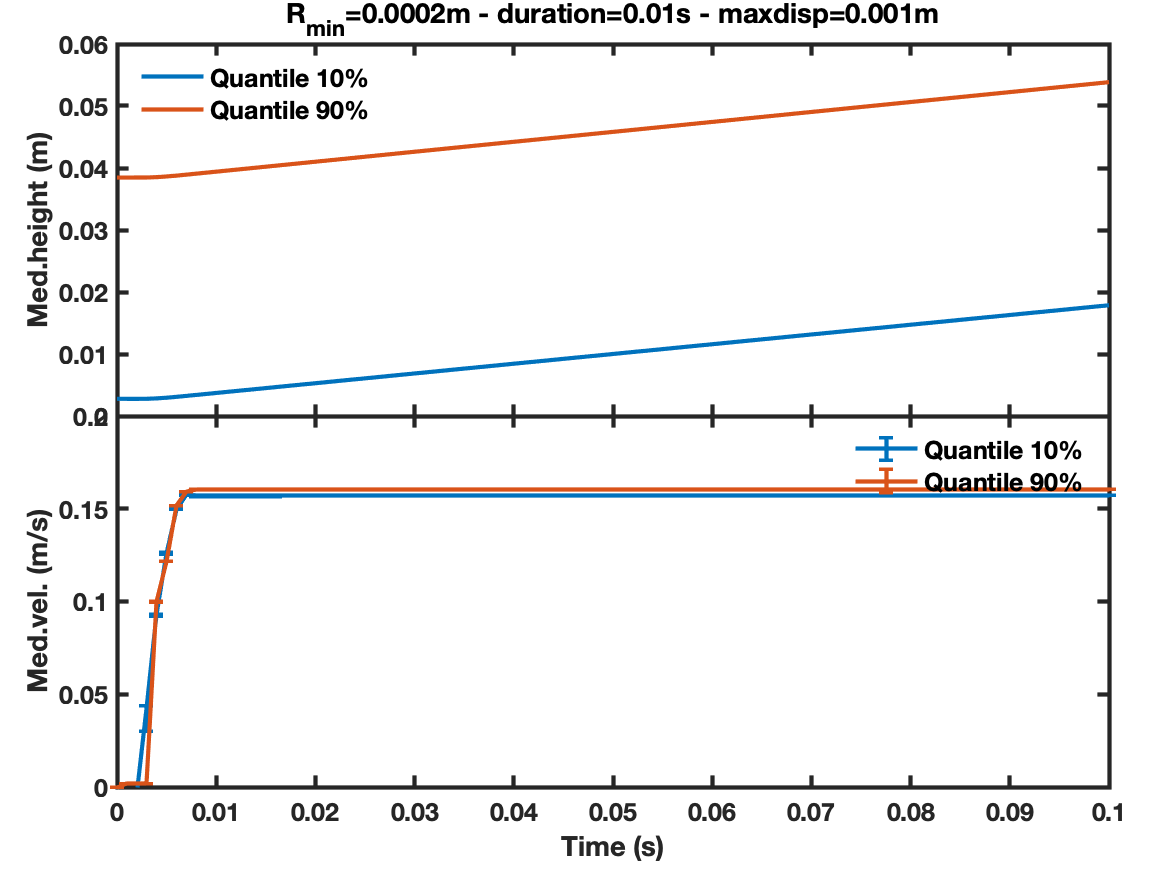}
    \includegraphics[width=1\linewidth]{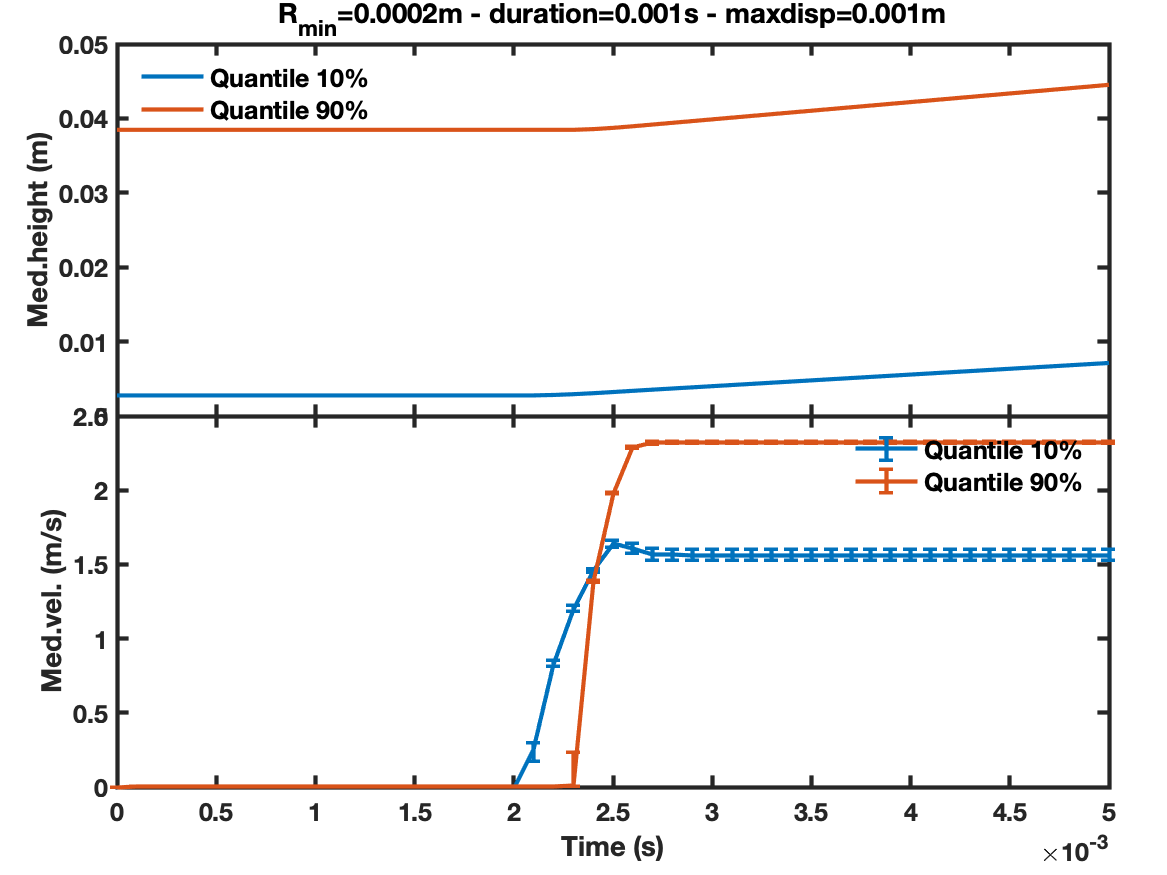}
    \caption{The median height and velocities of the bottom 10\% quantile and the top 90\% quantile of  particles ordered by height. We present results from two experiments of $\sin^2$ displacement, they have the same amplitude of  displacement,  $10^{-3}$ m, but different half-duration: \textit{a)} $10^{-2} \ s$ and \textit{b)} $10^{-3} \ s$, respectively.}
    \label{fig:evoldisp}
\end{figure}

In Fig. \ref{shaking}\textit{a} we present the initial status of the box with  particles at rest. After we apply the $\sin^2$ displacement to the floor, the particles glued with the floor move upwards, and they knock the particles above. This triggers  particle motion upwards, but due to mutual collisions, there are  differential velocity displacements. The particles on the upper layer move  faster than the lower ones, as  can be seen in Fig. \ref{fig:shaking}\textit{b}, at 5 milliseconds after the start. A vertical velocity gradient can be observed. To show the differences between the upper and lower layers, we compute the median height and velocities of the bottom 10\% quantile and the top 90\% quantile of  particles ordered by height. The evolution of these parameters is shown in Fig. \ref{fig:evoldisp}. We present results from two experiments of $\sin^2$ displacement, they have the same amplitude of  displacement,  $10^{-3} \ m$, but different half-duration: $10^{-2}$ s and $10^{-3} \ s$, respectively. In both cases, the displacement starts at $2\times10^{-3} \ s$ after the simulation starts, like the one in Fig. \ref{fig:floordisp}. In Fig. \ref{fig:evoldisp}, we observe that the bottom layer starts to move upwards, and afterwards the upper layer starts moving. When the pulse is fast (\textit{i.e.} half-duration of $10^{-3} \ s$), the difference between the velocity of the upper to the lower layer is wider; but, if the pulse is slow, the upper layer goes along the displacement of the lower layer.

\subsection{Summary of results of the shaking process}

The displacement  of the floor of a regolith layer of $mm$-size particles, in a hundredth of second (or faster), produces the lofting of the upper part of the layer, with particle ejection velocities comparable to the escape velocity of Dimorphos ($\sim 0.1 \ m/s$).

Since, in   Section \ref{resprop},  we observed that this kind of displacement would take place in a large area of the impacted hemisphere, we conclude that the lofted material will come from large distances from the impact point. Unlike the plume cone, in which  dust particles are generally moving radially from the impact point, the lofted material due to shaking will be distributed nearly in an isotropic way.

Let's consider there is a $10 \ cm$ regolith layer covering the surface of Dimorphos, similar to the layers observed in Eros \citep{richardson2005} and Itokawa \citep{michel2009}. Based on the results in Section \ref{resprop}, we estimate the area affected by the shaking mechanism of at least $\sim 1/4$ of the total surface area of Dimorphos (\textit{i.e.} $\sim 2\times10^4 \ m^2$). Assuming a particle density of $3000 \ kg/m^3$, and macroscopic porosity of 50\%, we get a total mass of the affected regolith layer of $\sim 3\times10^6 \ kg$. This amount of mass in $mm$ to $cm$-size dust particles could be ejected at velocities comparable to the escape velocity.

Using the point source impact scaling laws by \citet{housen2011}, \citet{cheng2022} and \citet{moreno2022} estimated a total ejected mass due to crater formation on Dimorphos of $\sim 4-5\times10^6$ kg (at velocities above $v_{esc,Dim}$), which is comparable to our value of lofted material. Using \textit{SPH} simulations, \citep{ferrari2022} made an estimate of the total ejected mass with velocities above the $v_{esc,Dim}$ of $\sim 5\times10^3 \times$ the mass of the impactor, \textit{i.e.} $\sim 2.8\times10^6 \ kg$; a similar value to the previous ones. The main difference is that the lofted material would have ejection velocities just above the escape velocities, while the material in the plume will move much faster. This difference  has important implications for the observability of the ejecta cloud, as we will show next.

Obviously, there are huge uncertainties in the above estimates, because we do not know the surface structure of Dimorphos, and we only have a rough estimate of the affected area, which depends, as   was mentioned   in Section \ref{resprop}, on many free parameters of   simulations. We just want to emphasize that the "cocoa effect" could be a relevant phenomena that cannot be disregarded.

\section{Ejection of particles and predictions for the observations} \label{observation}

The evolution of the particles ejected from Dimorphos's surface at different speeds was already studied by our group, and some results were presented in \citet{moreno2022}.

Two sets of observations of the aftermath of the DART experiment will be performed: \textit{i)} The observation of evolution of the impact itself, in  the first few minutes  during the LICIACube fly--by, with enough resolution to closely observe  Dimorphos' and Didymos' surfaces; \textit{ii)} The long-term monitoring from distant telescopes (from ground and space) of the dust coma and tail generated by the impact. We will analyze the detectability of the ejecta in these two sets of observations.

The evolution of the dust cloud generated by the impact has a strong dependence on the ejection speeds. 
The ejected material can be broadlyclassified into 3 different regimes: the very-high-speed ejecta, with  velocities from the surface over $v>10 \ m/s$; the high-speed ejecta, with $1 < v < 10 \ m/s$, and the low-speed ejecta, with $v<1 \ m/s$.

The very-high and high-speed ejecta are generated in the cratering event and from the crater area, and are ejected within the conical plume. The very-high speed ejecta will be observed by LICIACube in the first seconds after the impact; since, by the time of  close approach ($t_{cl} = +167 \ s$ after impact), the very-high speed ejecta will overtake Didymos. If these very--high speed ejecta ever exist, and depending on the geometry of the actual impact and the orientation of the plume with respect to Didymos, some dust particles could travel directly towards Didymos, producing some observable  impact flash  on the surface, similar to the impact flashes observed on the Earth's Moon surface; but it might be difficult to detect them due to the rate of image acquisition by LICIACube.

The high-speed ejecta will be seen by LICIACube slowly moving within the conical plume during the approaching and the receding phases. At $t_{cl}$, most of the high--speed material will still be within the Didymos  system, at distances of a few hundreds $m$ up to slightly over one $km$ from Dimorphos.

The visibility of the ejecta plume was discussed in detail by \citet{cheng2020} and \citet{cheng2022}. Hereby, we analyze the evolution and observability of the low-speed ejecta.

\subsection{The first seconds after impact and the observations by LICIACube}

Part of the low-speed ejecta will be formed in the cratering event and they will move within the conical plume. In addition, as we propose in the previous sections, the propagation of the impact-induced seismic wave will produce a strong shaking in a large area surrounding the impact point. Such a shaking  will trigger the occurrence of the "cocoa effect": the lofting at very low velocity of fine particles coming from a large surface area. Lofted particles will move away from Dimorphos surface at speeds comparable to the escape speed ($v_{esc,Dim} \approx 0.089 \ m/s$). Let's recall that the existence of  low--speed ejecta is supported by  observations of long--lasting dust tails in Active Asteroids.

Particles moving with velocities $v < v_{esc,Dim}$ will fall back after suborbital flight that could last minutes. Therefore, during the LICIACube fly-by there will be dust material just above the surface. Particles moving upwards with $v_{esc,Dim} < v < v_{esc,Sys}$ will be governed by the dynamics of the binary system. The seismic wave reaches a large fraction of the entire Dimorphos body in less than $2 s$, and they immediately produce shaking of the floor and   lofting of particles. At $t_{cl}$ lofted material at velocities a few times $v_{esc,Dim}$ will be a few $m$, up to a few tens of $m$, above the surface of the impacted hemisphere. If the impact--induced seismic wave ever reaches the antipodal hemisphere, lofted material could be observed over the entire surface.

\begin{figure}
    \centering
    \includegraphics[width=1\linewidth]{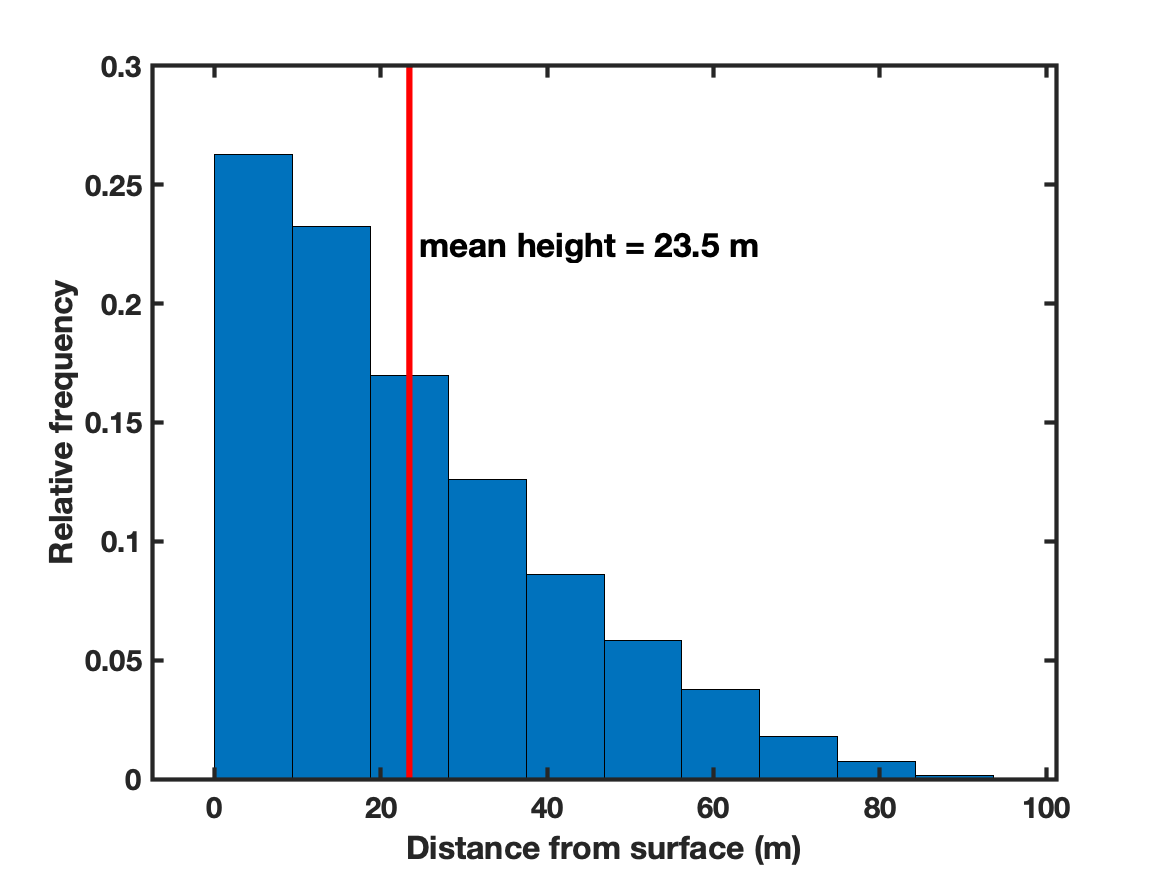}
   \caption{A histogram of the relative frequency of the distances from Dimorphos' surface of the 50,000 particles at $+100 \ s$ after impact. The distances were calculated with the model described in the text. The vertical red line corresponds to the mean height of $23.5 \ m$.}
    \label{fig:height}
\end{figure}

The Hill's radius of Dimorphos in the binary system is $\sim~171 \ m$, roughly $2\times$ Dimorphos radius. Particles leaving Dimorphos  surface at ejection velocities a few times $v_{esc,Dim}$ will move inside Dimorphos  Hill's radius in the first tens of seconds after the impact. Therefore, to study the early evolution of the ejecta cloud at low speed, we consider only the motion of  particles under the influence of Dimorphos  gravity field. Let's consider a Monte Carlo model for a set of particles leaving Dimorphos  surface. Ejection velocities are in a range spanning from the circular velocity at the surface ($v_{circ,Dim}=0.063 \ m/s$) up to $10\times v_{esc,Dim}$, with a triangular shaped distribution. The lower limit of   speed is such that particles can rise from the surface. Velocity vectors are pointing outwards with an isotropic distribution in the outgoing hemisphere. Particles with velocities $v$ from $v_{circ,Dim}$ up to $v_{esc,Dim}$ will get elliptical orbits, and lately fall back to Dimorphos. Particles with velocities $v > v_{esc,Dim}$ will move in  hyperbolic orbits away from Dimorphos, asymptotically approaching the velocity at infinity given by: $U = \sqrt{v^2 - v_{esc,Dim}^2}$. By solving the Kepler's equations, we compute the distance of the particles from Dimorphos' surface at a given time after impact (\textit{e.g.} $+100 \ s$). With the above range of ejection velocities,  particles will be at $100 \ s$ after impact at heights above the surface from $0$ up to $90 \ m$ (Fig. \ref{fig:height}), with a mean value of $23.5 \ m$. At the time of closest approach ($t_{cl} = +167 \ s$), the mean height of the cloud would be $42 \ m$.

Such dust particles at low height from the surface and spread over a large area would obscure the visibility of Dimorphos surface, producing a hazy effect, similar to the picture taken after the Mexicali earthquake (Fig. \ref{fig:mexicali}).

A precise estimate of the effect of this dust cloud on the observability from LICIACube is a great challenge, due to the several uncertainties in the model, like the total amount of ejected mass, the size range and size distribution of   particles, the extent of the area affected by the shaking mechanism, the velocity distribution of   particles, among others. Just to have an idea of the relevance of this effect, let's assume the following model: \textit{i}) a layer of particles above Dimorphos' surface with a given height $h$, following the results of the ejection model described above; \textit{ii}) a total ejected mass in low-speed ejecta of $M_{ej} \sim 3\times10^6 \ kg$; \textit{iii}) particles in a size range from $1 \ mm$ to $1 \ cm$; with a power-law size distribution similar to the one observed on the surface of asteroid Itokawa, with a differential mass index of $\beta=-3.2$ \citep{tancredi2015}; \textit{iv}) particle's density $\rho = 2170 \ kg/m^3$.

The amount by which each particle of radius $r$ reduces the radiant beam $I_0$ is $\pi r^2  k_e  I_0$, where $k_e$ is the extinction coefficient. Adding the contribution of the cloud of particles, we obtain the following expression for the optical depth:

\begin{equation}
\tau = \pi h \int_{0}^{\infty} k_e(r) r^2  n(r)  dr
\label{eq:tau}
\end{equation} 

where $n(r)$ is the number density distribution of particles. $n(r)~=~f(r)/V$, where $f(r)$ is the size distribution of  particles, and $V$ is the volume of the cloud over the impacted hemisphere: $V=\frac{2}{3} \pi \left( \left( r_{Dim}+h \right)^3-r_{Dim}^3\right)$. We are assuming that the ejected particles are coming mainly from the impacted hemisphere, which   happens to be the hemisphere seen by LICIACube in its approaching phase.

In this model, we are assuming that there is no mutual shadowing of the particles,  particles scatter  light individually and no effect of multiple scattering is considered. 

We assume a mass distribution:

\begin{equation}
  f(m)=\begin{cases}
    C m^\beta, & \text{if $m \in [a,b]$}.\\
    0, & \text{otherwise}.
  \end{cases}
\end{equation}

The parameter $C$ can be obtained from the equation of the total mass of the cloud:

\begin{equation}
M = \int_{0}^{\infty} m  f(m)  dm = \int_{a}^{b} m  C m^\beta  dm = \frac{C}{\left( \beta+2\right)} {b^{\beta+2}-a^{\beta+2}}
\label{eq:mass}
\end{equation} 

We then obtain:

\begin{equation}
    C=\frac{M \left( \beta+2\right)}{b^{\beta+2}-a^{\beta+2}}
\end{equation}

Replacing in eq. \ref{eq:tau} $f(r) dr = f(m) dm$, and integrating in the range $[a,b]$, we obtain:

\begin{equation}
    \tau = 2 \pi \frac{h}{V}  \left( \frac{3}{4 \pi \rho}\right)^{\frac{2}{3}}  \frac{C}{\beta+\frac{5}{3}} \left( b^{\beta+\frac{5}{3}}-a^{\beta+\frac{5}{3}} \right)
\end{equation}

Considering that the mean height of the cloud at $100 \ s$ after impact is $h=23 \ m$, we obtain an optical depth over the surface of Dimorphos of $\tau \sim 21$. Only $\exp(-\tau) =  7\times10^{-10}$ of the radiation coming from Dimorphos surface would be observed by LICIACube. At the time of the closest approach ($t_{cl} = +167 \ s$), the optical depth will reduce to $\tau \sim 17$, and the total brightness reduction will be by a factor of $3\times10^{-8}$.

If the amount of dust released by the shaking mechanism is reduced by a factor of 10 with respect to the previously assumed value (\textit{i.e.} $M_{ej}~\sim~3\times10^5$), the optical depth at $t=+100 \ s$ will be reduced by a simular factor to $\tau \sim 2$ and the brightness reduction factor to $\sim 0.12$; and at closest approach the numbers are $\tau \sim 1.7$ and the brightness reduction factor $\sim 0.18$.

In any if these scenarios, the radiation coming from Dimorphos  surface will be considerably absorbed and scattered by the dust cloud. 
In addition to the absorption effect, there would be a strong blurring effect that would make it difficult to detect  surface features. It would not be possible to clearly observe Dimorphos' surface, as it happened with the hills behind the dust cloud in the Mexicali earthquake.

\begin{figure}
    \centering
    \includegraphics[width=1\linewidth]{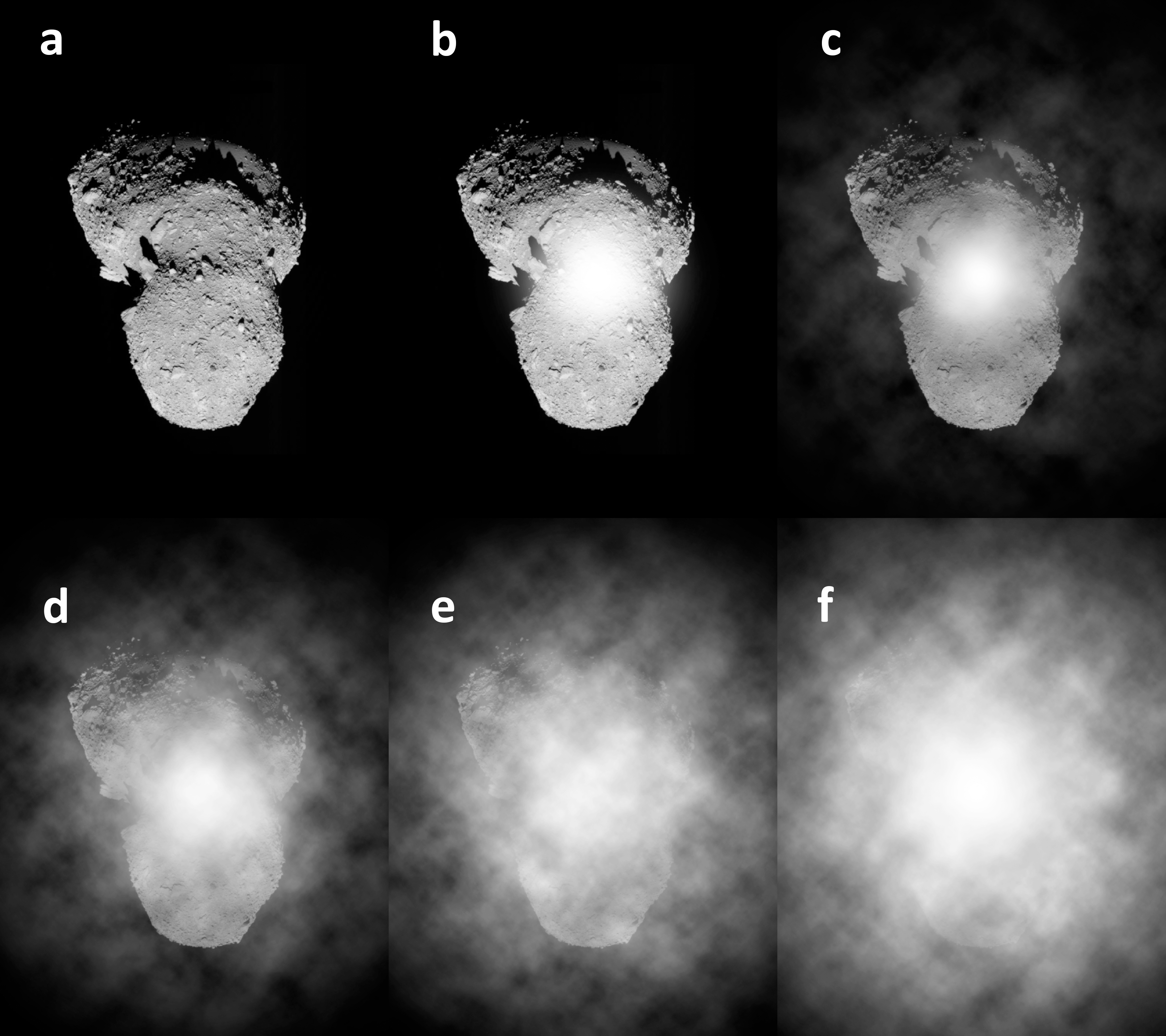}
    \caption{A set of artistic renders of the dust cloud, using as a model an image of Itokawa taken by Hayabusa. We assume that LICIACube's camera is pointing directly to the impact point. The sequence corresponds to an image at $\sim 100$ s after impact and alternative scenarios depending on the amount of material ejected at low-speeds. The sequence starts with \textit{a)} the bare body, \textit{b)} the brightening and the conical plume seen from the axis direction, with negligible contribution from the low-speed ejecta, \textit{c)} to \textit{f)} an image sequence with an increasing contribution of low speed ejecta from a wide area of the impacted hemisphere.}
    \label{fig:itokawa}
\end{figure}

As an illustration of the different scenarios that LICIACube could face, we produce a sequence of artist's render of the dust cloud using an image of Itokawa taken by Hayabusa as a model (Fig \ref{fig:itokawa}). We assume that LICIACube's camera is pointing directly to the impact point. The sequence corresponds to a snapshot taken at $\sim 100 \ s$ after impact and alternative scenarios depending on the amount of material ejected at low-speeds. The sequence starts with \textit{a)} the bare body, \textit{b)} the brightening and the conical plume seen from the direction of the cone axis, with negligible contribution from the low-speed ejecta, \textit{c)} to \textit{f)} an image sequence with an increasing contribution of low velocity ejecta from a wide area of the impacted hemisphere.   

This issue will not affect the observation of Didymos from LICIACube, unless a very particular configuration of a quasi-mutual eclipse happens, since the dust cloud will be very close to Dimorphos' surface.

\subsection{The formation of the coma and the tail, and the observations from distant telescopes}

In our previous work \citep{moreno2022}, we studied the evolution of the brightness of the coma and the tail as a function of  ejection speed. 
We performed simulations with a wide range of possible ejection velocities, from Dimorphos escape velocity of $0.09 \ m/s$ up to $800 \ m/s$. A nominal value of the total ejected mass of $5\times10^6 \ kg$ was used for the calculations; although, we also presented some results with a larger amount of ejected mass.

A Monte Carlo approach was used to model the evolution of ${\mu}m$ to $cm$ dust particles under the gravity forces by Dimorphos, Didymos and the Sun, and  solar radiation pressure, the only non-gravitational force acting on particles that we considered in the model. The Poynting--Robertson drag on those particles is only important in very long-term dynamics, and was therefore neglected.   Particles follow a differential power-law  size distribution with  exponent  $-3.5$ in the range from $1 \ {\mu}m$ to $1 \ cm$. We propagated the orbits of    ejected particles by integrating their equation of motion, and the results were used to study the evolution of the coma and the tail brightness.

Note that in this model we considered particles down to $1\ {\mu}m$ in size. Those very small particles will be strongly affected by the solar radiation pressure, even at the early stages of ejection. For example, a $1\ {\mu}m$ particle released from the surface at $0.1 \ m/s$ will be accelerated up to a velocity of $5$ m/s, $2 \ h$  after the impact; while a $100\ {\mu} m$ particle will have a velocity of $0.2 \ m/s$ after the same time. Therefore, the   smallest particles will disperse quickly, and their contribution to the coma and the long-lasting tail will be small, as it has been observed in the modelling of Active Asteroids \citep{moreno2021}.

The model computed the asteroid tail surface brightness and the integrated brightness of the coma at a given date as seen from the Earth. The nucleus brightness of Didymos was also included in the simulated images.

For low-speed ejecta, up to a several times Dimorphos escape velocity, \citet{moreno2022} predicted an increase of brightness of $\sim 3$~ mag right after the impact, and decay to pre-impact levels some 10 days after \citep[see Fig. 7 in][]{moreno2022}. However, if most of the ejecta is released at speeds of the order of $\ge 100 \ m/s$, the observability of the event would reduce to a very short time span, of the order of one day or shorter.

\begin{figure}
    \centering
    \includegraphics[width=1\linewidth]{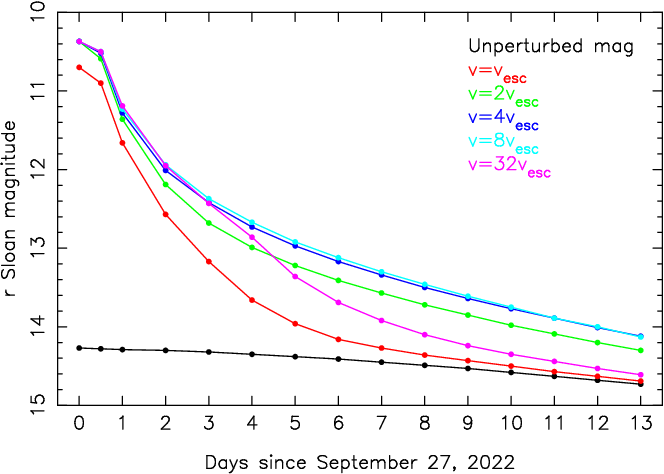}
    \caption{Calculated \textit{r}-Sloan magnitudes as a function of date and ejection velocities, for slow ejection speeds, compared to the unperturbed system magnitude. See \citet{moreno2022} for the details of the model.}
    \label{fig:brightness}
\end{figure}

To discuss in more detail the effects of the low-speed ejecta, we extend Fig. 7 of \citet{moreno2022} to include several curves for the low-speed cases (Fig. \ref{fig:brightness}), and extending the calculations to show results at still shorter times relative to the impact time. In this new plot, the first data points are at Sept. 27 00:00 UT, i.e., only $46 \ min$ after impact. During the first few hours after impact the increase of brightness relative to the unperturbed system could reach up to $\sim 4 \ mag$ for ejecta speeds between 2 and $32 \times v_{esc,Dim}$.

Let's consider the case of particles with ejection speeds $4\times v_{esc}$. Immediately after the impact, the object will increase its brightness by $\sim 4 \ mag$ with respect to the bare  Didymos system; 5 days after the impact, the integrated magnitude of the coma will still be $1.4 \ mag$ brighter than the nucleus; and 10 days after the impact, the difference will be $0.8 \ mag$. In the latter case, the total flux will be twice the nucleus  flux; therefore, half of the total brightness would come from the bare system and the other half from the surrounding coma. According to previous studies \citep{pravec2006}, the magnitude drop  due to   mutual events is $\sim 0.1 \ mag$. Because of the large contribution of the dust coma to the total brightness in the first few days after impact, the small dips associated with mutual events would be masked and difficult to detect.

The comma flux scales approximately with the total particle area and, consequently, with the total mass ejected. Therefore, a factor of $10\times$ increase (decrease) on the amount of ejected mass would produce a $\sim 2.5 \log_{10} 10 = 2.5 \ mag$ upward (downward) shift of the curves in Fig. \ref{fig:brightness} \citep[see Fig. 9 of ][]{moreno2022}. 

Monitoring the evolution of  total brightness in the first 2-3 days after impact can be used to estimate when the contribution of the coma will be negligible. The observed magnitudes can be located in the plot of Fig. \ref{fig:brightness}, and the curve that best matches observations can be used to extrapolate for later days.

\section{Conclusions and discussion} \label{discussion}

The DART impact experiment on the Didymos satellite, Dimorphos will take place on the 26$^{th}$ of September, 2022 at 23:14 UT. The impact will produce a crater and ejected material at high-speed in a conical plume. In addition, impact generated seismic waves will propagate inside Dimorphos. The attenuation of those waves depends on the mechanical properties of the material, which are largely unknown. Using some realistic mechanical properties of rocky material in two different types of DEM simulations, we studied the propagation of the waves at a large distance from the impact point on the surface. We worked out that, from a few tenths of a second after impact up to $2 \ s$, the particles located several tens of degrees from the impact point may be shaken due to the incoming seismic wave. If a thin regolith layer located on the surface is shaken, the top of the layer can be ejected at velocities comparable to the escape velocity from the Dimorphos surface. We call this phenomena the "cocoa effect". We estimate that an area on the order of at least $1/4$ of the total surface of Dimorphos could be shaken and several thousand tons of dust could be released at low velocities (less than $1 \ m/s$). The motion of the ejected dust particles will be affected by the gravitational forces of mainly Dimorphos, Didymos, and the Sun, as well as the solar radiation pressure.

As a consequence of such sequence of events, we anticipate the following potentially observable effects: \textit{i)}   generation of a cloud of small particles that would produce a hazy or fuzzy appearance of Dimorphos  surface, detectable by LICIACube; \textit{ii)}  brightness increases of the binary system due to enhancement of the cross section produced by the cloud of particles; \textit{iii)} generation of a dust trail, similar to those observed in Active Asteroids, which can last  several weeks after impact. 

A prediction of the observability of these effects strongly depends on the amount and size distribution of  ejected particles, which are largely unknown. The monitoring of the Didymos system in the hours and days after the impact can be used to discuss the relevance of the different ejection mechanisms; and, by an inversion analysis, estimates of  some elastic and structural parameters of Dimorphos can be obtained.

It is worth saying that the presence of  low-speed ejecta might not have relevant consequences for the key objective of the DART mission, \textit{i.e.} get an estimate of the efficiency of linear momentum transfer from the projectile to the target. Assuming a total ejected mass at low-speed  of $\sim 3\times10^6 \ kg$, the kinetic energy carried by the dust cloud at velocities of $\sim 0.1 \ m/s$ would be $<10^{-4}\times$ the kinetic energy of the DART spacecraft. The sum of the modulus of the linear momentum of the released particles would be $~0.1\times$ the linear momentum provided by the DART spacecraft. However,  the net effect depends on the vector addition of the individual linear momenta of particles, and therefore such effect will be largely less than 0.1.

Finally, we emphasize that the DART experiment can be considered as the production of an artificial Active Asteroid, with an extensive set of follow-up activities at the moment of impact and beyond. Therefore, the DART experiment will also provide --as an additional side product-- useful information for understanding the physics of the generation of Active Asteroids.

\section*{Acknowledgements}


GT and BD acknowledge financial support from project FCE-1-2019-1-156451 of the Agencia Nacional de Investigaci\'on e Innovaci\'on ANII (Uruguay). The ESyS simulations were run in the ClusterUY - Centro Nacional de Supercomputaci\'on de Uruguay (\textsc{}{https://www.cluster.uy/}). ACB and PYL acknowledge funding by the SU-SPACE-23-SEC-2019 EC-H2020 NEO-MAPP project (GA 870377). ACB also acknowledges funding by the Spanish Ministerio de Ciencia e Innovaci\'on RTI2018-099464-B-I00 project.
FM acknowledges financial support from the State Agency for Research of the Spanish MCIU through the "Center of Excellence Severo Ochoa" award to the Instituto de Astrof\'isica de Andaluc\'ia (SEV-2017-0709). FM also acknowledges financial support from the Spanish Plan Nacional de Astronom\'ia y Astrof\'isica LEONIDAS project RTI2018-095330-B-100, and project P18-RT-1854 from Junta de Andaluc\'ia.  
\section*{Data Availability}


The plots included in this article were generated with data coming out of the simulations. We can provide to the interested readers with the initial conditions of the simulations in more detail than the ones described in the text, as well as the outcomes of the simulations. Please, contact the corresponding author.



\bibliographystyle{mnras}
\bibliography{refs}








\bsp	
\label{lastpage}
\end{document}